\documentclass[preprint,12pt]{elsarticle}
\usepackage{amsmath}
\usepackage{subfigure}
\usepackage{amssymb}
\usepackage{bm}
\usepackage{booktabs}
\usepackage{graphicx,graphics,dcolumn,booktabs,bm}
\usepackage{epsfig}
\usepackage{epstopdf}   
\usepackage{slashed}  
\usepackage{multirow}
\usepackage{color}

\journal{Nuclear Physics A}

\allowdisplaybreaks

\begin{document}

\begin{frontmatter}
\title{Hidden-charm molecular pentaquarks and their charm-strange partners}
\author[1,2]{Rui Chen}
\author[1,2]{Xiang Liu}
\author[3,4,5]{Shi-Lin Zhu}
\address[1]{
School of Physical Science and Technology, Lanzhou University,
Lanzhou 730000, China}

\address[2]{Research Center for Hadron and CSR Physics, Lanzhou University and Institute of Modern Physics of CAS, Lanzhou 730000, China}

\address[3]{School of Physics and State Key Laboratory of Nuclear Physics
and Technology, Peking University, Beijing 100871, China}

\address[4]{Collaborative Innovation Center of Quantum Matter, Beijing
100871, China}

\address[5]{Center of High Energy Physics, Peking University, Beijing
100871, China }

\begin{abstract}
In the framework of one-pion-exchange (OPE) model, we study the
hidden-charm and charm-strange molecular pentaquark systems composed
of a heavy baryon $(\Sigma_c, \Sigma_c^*)$ and a vector meson
$(\bar{K}^*, \bar{D}^*)$, where the S-D mixing effect is considered
in our calculation. Our result shows that the $\Sigma_c\bar{D}^*$
molecular state with $(I=1/2,J^P=3/2^-)$ and the
$\Sigma_c^*\bar{D}^*$ molecular state with $(I=1/2,J^P=5/2^-)$ exist
in the mass range of the observed $P_c(4380)$ and $P_c(4450)$,
respectively. Moreover, we predict two other hidden-charm molecular
pentaquarks with configurations $\Sigma_c\bar{D}^*$ $(I=3/2,
J^P=1/2^-)$ and $\Sigma_c^*\bar{D}^*$ $(I=3/2, J^P=1/2^-)$ and two
charm-strange molecular pentaquarks $P_{cs}(3340)$ and
$P_{cs}(3400)$ corresponding to the $\Sigma_c\bar{K}^*$
configuration with $(I=1/2, J^P=3/2^-)$ and the
$\Sigma_c^*\bar{K}^*$ configuration with
$(I=1/2, J^P=5/2^-)$, respectively. Additionally, we
also predict some hidden-bottom $\Sigma_b^{(*)}B^*$ and $B_c$-like
$\Sigma_c^{(*)}B^*/\Sigma_b^{(*)}\bar{D}^*$ pentaquarks.

\end{abstract}

\begin{keyword}

Molecular pentaquark state, Exotic state, One pion exchange, S-D
mixing

\PACS 12.39.Pn\sep 14.20.Pt
\end{keyword}

\end{frontmatter}

\section{Introduction}\label{sec1}

One of the most important research topics of hadron physics is the
search for the exotic states like the glueballs, hybrid mesons, and
multiquark states. In the past decade, the experimental observations
of many charmonium-like states have inspired extensive study of
exotic hadronic states (see the recent review in Ref.
\cite{review}). Especially, the recent observation of $P_c(4380)$
and $P_c(4450)$ states by the LHCb Collaboration \cite{Aaij:2015tga}
in the $\Lambda_b^0\rightarrow K^-J/\psi p$ process has aroused the
theorists' strong interest in the hidden-charm pentaquark states.
The resonance parameters of the two $P_c$ states are
\cite{Aaij:2015tga}
\begin{eqnarray*}\begin{array}{ll}
M_{P_c(4380)}=4380\pm 8\pm 29\, \text{MeV},
&\Gamma_{P_c(4380)}=205\pm 18\pm 86\, \text{MeV}, \\
M_{P_c(4450)}=4449.8\pm 1.7\pm 2.5 \,\text{MeV},
&\Gamma_{P_c(4450)}=39\pm 5\pm 19 \,\text{MeV}.\end{array}
\end{eqnarray*}
The preferred spin-parity quantum numbers of of $P_c(4380)$ and
$P_c(4450)$ are $3/2^{\mp}$ and $5/2^\pm$, respectively. Before the
observation of the two $P_c$ states, hidden-charm pentaquarks have
been predicted in Refs.
\cite{Yang:2011wz,Wu:2010jy,Wu:2010vk,Wang:2011rga,Garcia-Recio:2013gaa,
Xiao:2013yca,Yuan:2012wz,Uchino:2015uha}.

Until now, theoretical interpretations of $P_c(4380)$ and
$P_c(4450)$ include the molecular pentaquark states
\cite{Chen:2015loa,Chen:2015moa,Karliner:2015ina,Roca:2015dva,
Mironov:2015ica,He:2015cea,Meissner:2015mza,Burns:2015dwa,Huang:2015uda},
the diquark-diquark-antiquark pentaquark
\cite{Maiani:2015vwa,Anisovich:2015cia,Li:2015gta,
Ghosh:2015ksa,Wang:2015epa,Anisovich:2015zqa}, the diquark-triquark
pentaquark \cite{Lebed:2015tna,Zhu:2015bba}, the re-scattering
effect \cite{Guo:2015umn,Liu:2015fea,Mikhasenko:2015vca}, the
topological soliton model \cite{Scoccola:2015nia}, etc. With the
molecular pentaquark assignment, the mass of $P_c(4450)$
and $P_c(4380)$ can be understood quite naturally
\cite{Chen:2015loa}.

\begin{figure}[!htbp]
  \centering
  \includegraphics[width=4in]{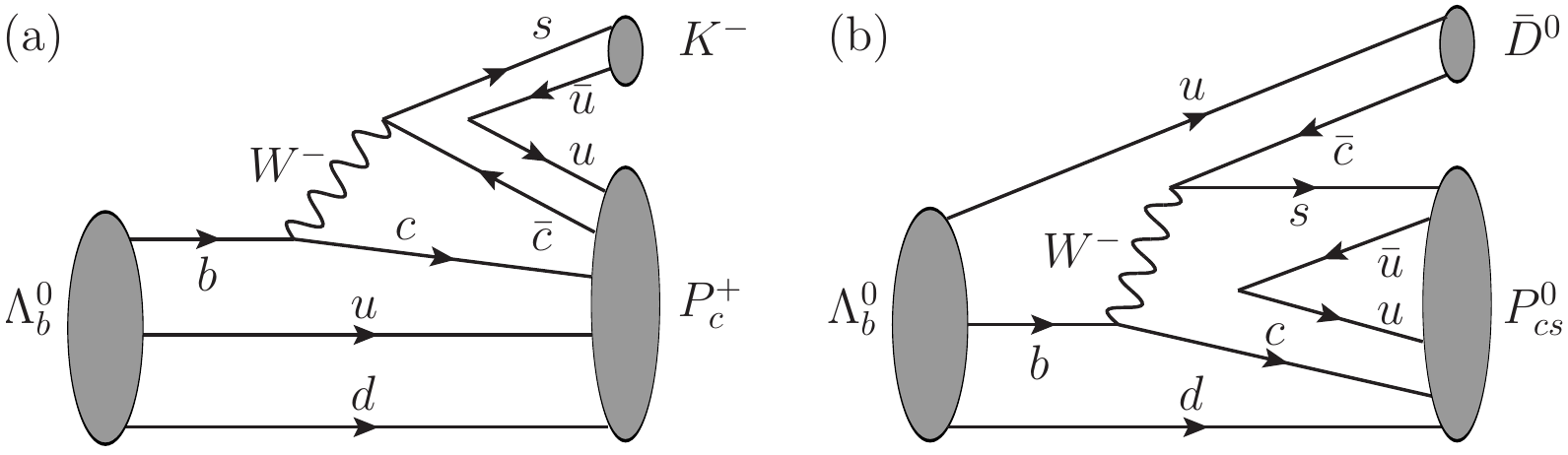}\\
  \caption{Feynman diagrams for $\Lambda_b^0\to K^-P_c^+$ and $\Lambda_b^0\to \bar{D}^0 P_{cs}^0$.}\label{pcs}
\end{figure}

The tensor force mixes the S-D wave and plays a crucial role in the
case of the deuteron. This S-D mixing effect was not included in
Ref. \cite{Chen:2015loa}. In the present work, we consider the S-D
mixing effect to reexamine the hidden-charm
$\Sigma_c(2455)\bar{D}^*$ and $\Sigma_c^*(2520)\bar{D}^*$ systems,
which is an extension of Ref. \cite{Chen:2015loa}.

As shown in Fig. \ref{pcs}, two $P_c(4380)$ and $P_c(4450)$ states
are produced via Fig. \ref{pcs} (a). However, there exists another
combination of quarks in the final state in Fig. \ref{pcs} (b),
where the charm-strange pentaquarks (the $P_{cs}$ states) may also
be produced. Compared with Fig. \ref{pcs} (a), Fig. \ref{pcs} (b) is
not suppressed, which stimulates our interest in the charm-strange
molecular pentaquarks composed of either $\Sigma_c\bar{K}^*$ or
$\Sigma_c^*\bar{K}^*$.

We also study the hidden-bottom $\Sigma_bB^*$ and $\Sigma_b^*B^*$
molecular systems and $B_c$-like $\Sigma_cB^*/\Sigma_b\bar{D}^*$ and
$\Sigma_c^*B^*/\Sigma_b^*\bar{D}^*$ molecular systems, which are the
partners of the hidden-charm molecular pentaquarks.

This paper is organized as follows. After the introduction, we
present the deduction of the OPE potentials in Section \ref{sec2}
and numerical results in Section \ref{sec3}. The last section is a
short summary.

\section{The effective potential}\label{sec2}

In the following, we derive the OPE effective potentials of the
hidden-charm $\Sigma_c\bar{D}^*$ and $\Sigma_c^*\bar{D}^*$ molecular
systems, and charmed-strange $\Sigma_c\bar{K}^*$ and
$\Sigma_c^*\bar{K}^*$ molecular systems.

\subsection{Wave function}\label{fuction}

In this subsection, we construct the wave functions of the
hidden-charm $\Sigma_c\bar{D}^*$ and $\Sigma_c^*\bar{D}^*$ molecular
systems and charmed-strange $\Sigma_c\bar{K}^*$ and
$\Sigma_c^*\bar{K}^*$ molecular systems. For a system composed of
two colorless hadrons, its total wave function is expressed as
\begin{eqnarray}
\left|\Psi \right\rangle &=&\left |\frac{\phi(r)}{r}\right
\rangle\otimes\left|{}^{2S+1}L_{J}\right\rangle\otimes|I,I_3\rangle,
\end{eqnarray}
where $|\phi(r)/r\rangle$, $|{}^{2S+1}L_{J}\rangle$ and
$|I,I_3\rangle$ denote the radial, spin-orbit and flavor wave
functions, respectively. The radial wave function will be obtained
via numerical calculation in Sec. \ref{sec3}. Since the S-D mixing
is taken into account, the orbital quantum number $L=0$ or $L=2$.
The spin $S=1/2$ and $3/2$ for the $\Sigma_c\bar{M}^*$
configuration, and $S=1/2$, $3/2$ and $5/2$ for the
$\Sigma_c^*\bar{M}^*$ configuration. Then, the spin-orbit wave
function $|{}^{2S+1}L_J\rangle$ can be written as
\begin{eqnarray}
\begin{array}{ccccc}
J=\frac{1}{2}:    &|{}^2\mathbb{S}_{\frac{1}{2}}\rangle, &|{}^4\mathbb{D}_{\frac{1}{2}}\rangle,\\
J=\frac{3}{2}:   &|{}^4\mathbb{S}_{\frac{3}{2}}\rangle,   &|{}^2\mathbb{D}_{\frac{3}{2}}\rangle,    &|{}^4\mathbb{D}_{\frac{3}{2}}\rangle,\\
J=\frac{5}{2}:   &|{}^6\mathbb{S}_{\frac{5}{2}}\rangle,
&|{}^2\mathbb{D}_{\frac{5}{2}}\rangle,
&|{}^4\mathbb{D}_{\frac{5}{2}}\rangle,
&|{}^6\mathbb{D}_{\frac{5}{2}}\rangle.
\end{array}
\end{eqnarray}
The general expressions of the spin-orbit wave function
$|{}^{2S+1}L_J\rangle$ for the $\Sigma_c\bar{M}^*$ and
$\Sigma_c^*\bar{M}^*$ systems are
\begin{eqnarray}
\Sigma_c\bar{M}^*:\quad \left|{}^{2S+1}L_{J}\right\rangle &=&
\sum_{m,m',m_Sm_L}C^{S,m_S}_{\frac{1}{2}m,1m'}C^{J,M}_{Sm_S,Lm_L}
          \chi_{\frac{1}{2}m}\epsilon^{m'}|Y_{L,m_L}\rangle,\nonumber\\
\Sigma_c^*\bar{M}^*:\quad \left|{}^{2S+1}L_{J}\right\rangle &=&
\sum_{m,m',m_Sm_L}C^{S,m_S}_{\frac{3}{2}m,1m'}C^{J,M}_{Sm_S,Lm_L}
          \Phi_{\frac{3}{2}m}\epsilon^{m'}|Y_{L,m_L}\rangle.\nonumber
\end{eqnarray}
Here, $\chi_{\frac{1}{2}m}$ is the spin wave function and $Y_{L,m_L}$
is the spherical harmonics function. $C^{J,M}_{Sm_S,Lm_L}$,
$C^{S,m_S}_{\frac{1}{2}m,1m'}$ and $C^{S,m_S}_{\frac{3}{2}m,1m'}$
are the Clebsch-Gordan coefficients. The polarization vector for the
$\bar{M}^*$ vector meson is defined as
$\epsilon_{\pm}^{m}=\mp\frac{1}{\sqrt{2}}\left(\epsilon_x^{m}{\pm}i\epsilon_y^{m}\right)$
and $\epsilon_0^{m}=\epsilon_z^{m}$, which satisfy $\epsilon_{\pm1}
= \frac{1}{\sqrt{2}}\left(0,\pm1,i,0\right)$ and $\epsilon_{0} =
\left(0,0,0,-1\right)$. The polarization tensor
$\Phi_{\frac{3}{2}m}$ for $\Sigma_c^*$ is constructed as
$\Phi_{\frac{3}{2}m}=\sum_{m_1,m_2}\langle
\frac{1}{2},m_1;1,m_2|\frac{3}{2},m\rangle\chi_{\frac{1}{2},m_1}\epsilon^{m_2}$.

{For the $\Sigma_c\bar{M}^*$ and $\Sigma_c^*\bar{M}^*$
systems, the flavor wave function $|I,I_3\rangle$ can be grouped
into two categories since their isospin can be classified into
either $1/2$ or $3/2$,
i.e.,\begin{eqnarray}&&\left\{\begin{array}{c}
\left|\frac{1}{2},\frac{1}{2}\right\rangle =
     \sqrt{\frac{2}{3}}\left|\Sigma_c^{(*)++}{M}^{*-}\right\rangle
     -\frac{1}{\sqrt{3}}\left|\Sigma_c^{(*)+}\bar{M}^{*0}\right\rangle\\
\left|\frac{1}{2},-\frac{1}{2}\right\rangle =
     \frac{1}{\sqrt{3}}\left|\Sigma_c^{(*)+}{M}^{*-}\right\rangle
     -\sqrt{\frac{2}{3}}\left|\Sigma_c^{(*)0}\bar{M}^{*0}\right\rangle
     \end{array}\right.,\\
&&\left\{\begin{array}{l}
\left|\frac{3}{2},\frac{3}{2}\right\rangle = \left|\Sigma_c^{(*)++}\bar{M}^{*0}\right\rangle\\
\left|\frac{3}{2},\frac{1}{2}\right\rangle =
     \frac{1}{\sqrt{3}}\left|\Sigma_c^{(*)++}{M}^{*-}\right\rangle
     +\sqrt{\frac{2}{3}}\left|\Sigma_c^{(*)+}\bar{M}^{*0}\right\rangle\\
\left|\frac{3}{2},-\frac{1}{2}\right\rangle =
     \sqrt{\frac{2}{3}}\left|\Sigma_c^{(*)+}{M}^{*-}\right\rangle
     +\frac{1}{\sqrt{3}}\left|\Sigma_c^{(*)0}\bar{M}^{*0}\right\rangle\\
\left|\frac{3}{2},-\frac{3}{2}\right\rangle =
     \left|\Sigma_c^{(*)0}{M}^{*-}\right\rangle
     \end{array}\right.,
\end{eqnarray}
where $\bar{M}^*$ is defined as $\bar{M}^*=\left(K^{*-},-\bar{K}^{*0}\right)^T$ or $\bar{M}^*=\left(\bar{D}^{*0},{D}^{*-}\right)^T$.}

\subsection{Lagrangians} \label{lag}

In the derivation of the effective OPE potentials, we need the
effective Lagrangain
\begin{eqnarray} \mathcal{L}_{\mathbb{P}}&=&
       ig\text{Tr}\left[\bar{H_a}^{(\bar{Q})}\gamma^{\mu}A^{\mu}_{ab}
       \gamma_5H_b^{(\bar{Q})}\right],
     \label{lag01}\\
\mathcal{L}_{\mathcal{S}} &=&
-\frac{3}{2}g_1\varepsilon^{\mu\nu\lambda\kappa}v_{\kappa}\text{Tr}
      \left[\bar{\mathcal{S}}_{\mu}A_{\nu}\mathcal{S}_{\lambda}\right],\label{lag02}
\end{eqnarray}
which was constructed by considering the heavy quark limit and
chiral symmetry
\cite{Yan:1992gz,Wise:1992hn,Burdman:1992gh,Casalbuoni:1996pg,Falk:1992cx,Liu:2011xc}.
Here, $H_a^{(\bar{Q})}$ stands for a multiplet field composed of the
pseudoscalar meson $P^{(\bar{Q})}=(\bar{D}^0,D^-)^T$ and vector
meson $P^{*(\bar{Q})}=(\bar{D}^{*0}, D^{*-})^T$ in Eq. (\ref{hh}).
Its conjugate field satisfies
$\bar{H}_a^{(\bar{Q})}=\gamma_0H_a^{(\bar{Q})\dag}\gamma_0$.
$\mathcal{S}_{\mu}$ is the superfield composed of Dirac spinor
fields $\mathcal{B}_6$ with $J^P=1/2^+$ and $\mathcal{B}^*_6$ with
$J^P=3/2^+$ in the $6_F$ flavor representation. The expressions for
$H_a^{(\bar{Q})}$ and $\mathcal{S}_{\mu}$ read as
\begin{eqnarray}
H_a^{(\bar{Q})} &=& [P_a^{*(\bar{Q})\mu}\gamma_{\mu}-P_a^{(\bar{Q})}\gamma_5]\frac{1-\rlap\slash v}{2},\label{hh}\\
\mathcal{S}_{\mu} &=&
-\sqrt{\frac{1}{3}}(\gamma_{\mu}+v_{\mu})\gamma^5\mathcal{B}_6
       +\mathcal{B}_{6\mu}^*.\label{ss}
\end{eqnarray}
Here, $v=(1,\vec{0})$ is the four velocity under the non-relativistic
approximation. $A_{\mu} =
\frac{1}{2}(\xi^{\dag}\partial_{\mu}\xi-\xi\partial_{\mu}\xi^{\dag})$
is the axial current, where $\xi=\text{exp}(i\mathbb{P}/f_{\pi})$ and
$f_{\pi}=132$ MeV is the pion decay constant. The matrices for
$\mathbb{P}$, $\mathcal{B}_6$ and $\mathcal{B}_{6}^*$ are
\begin{eqnarray}
\mathbb{P} &=& \left(\begin{array}{cc}
\frac{\pi^0}{\sqrt{2}} &\pi^+\\
\pi^- &-\frac{\pi^0}{\sqrt{2}}
\end{array}\right),
\mathcal{B}_6 = \left(\begin{array}{cc}
         \Sigma_c^{++}              &\frac{\Sigma_c^{+}}{\sqrt{2}}\\
         \frac{\Sigma_c^{+}}{\sqrt{2}}      &\Sigma_c^{0}
\end{array}\right),
\mathcal{B}_6^* = \left(\begin{array}{cc}
         \Sigma_c^{*++}              &\frac{\Sigma_c^{*+}}{\sqrt{2}}\\
         \frac{\Sigma_c^{*+}}{\sqrt{2}}      &\Sigma_c^{*0}
\end{array}\right).\nonumber
\end{eqnarray}

Substituting Eq. (\ref{hh})-(\ref{ss}) into Eq.
(\ref{lag01})-(\ref{lag02}), we obtain
\begin{eqnarray}
\mathcal{L}_{\bar{P}^*\bar{P}^*\mathbb{P}} &=&
           i\frac{2g}{f_{\pi}}v^{\alpha}\varepsilon_{\alpha\mu\nu\lambda}
           \bar{P}_{a}^{*\mu\dag}\bar{P}_{b}^{*\lambda}\partial^{\nu}\mathbb{P}_{ab},\\
\mathcal{L}_{\mathcal{B}_6\mathcal{B}_6\mathbb{P}} &=&
      i\frac{g_1}{2f_{\pi}}\varepsilon^{\mu\nu\lambda\kappa}v_{\kappa}
      \text{Tr}\left[\bar{\mathcal{B}_6}\gamma_{\mu}\gamma_{\lambda}
      \partial_{\nu}\mathbb{P}\mathcal{B}_6\right],\\
\mathcal{L}_{\mathcal{B}_6^*\mathcal{B}_6^*\mathbb{P}} &=&
      -i\frac{3g_1}{2f_{\pi}}\varepsilon^{\mu\nu\lambda\kappa}v_{\kappa}
      \text{Tr}\left[\bar{\mathcal{B}}_{6\mu}^{*}\partial_{\nu}\mathbb{P}
      \mathcal{B}_{6\nu}^*\right].
\end{eqnarray}
Similar to the treatment in Ref. \cite{Liu:2008xz}, the pionic
coupling constant in Eq. (\ref{lag01}) is determined as $g=0.59\pm
0.07\pm 0.01$ from the $D^*$ decay width. The coupling constant
$g_1$ in Eq. (\ref{lag02}) is fixed as $g_1= 0.94$
\cite{Liu:2011xc}.

By gauging the Wess-Zumino term, the vector-vector-pseudoscalar
coupling was \cite{Chen:2011cj,Oh:2000qr},
\begin{eqnarray}
\mathcal {L}_{\bar{K}^*\bar{K}^*\mathbb{P}} &=& -\sqrt{2}g_{\pi
\bar{K}^*\bar{K}^*}\varepsilon^{\mu\nu\rho\sigma}
        \partial_{\rho}{\bar{K}^{*\dag}_{\sigma}}\mathbb{P}
        \partial_{\mu}\bar{K}^*_{\nu},
\end{eqnarray}
where $\bar{K}^*$ stands for $\bar{K}^*=(K^{*-}, \bar{K}^{*0})^T$.
The coupling constant $g_{\pi \bar{K}^*\bar{K}^*}$ is determined as
$g_{\pi \bar{K}^*\bar{K}^*} = \frac{g_2^2N_c}{64\pi^2f_{\pi}}$ with
$N_c=3$ and $g_2=13.7$.

We also need the normalization relations for the vector meson $M^*$,
baryon $\mathcal{B}_6$ and $\mathcal{B}_6^{*\mu}$
\begin{eqnarray*}
\langle0|\bar{M}^*_{\mu}|\bar{Q}q(1^-)\rangle &=&
\epsilon_{\mu}\sqrt{M_{\bar{M}^*}}, \\
\langle0|\mathcal{B}_6|Qqq(1/2^+)\rangle &=&
\sqrt{2M_{\mathcal{B}_6}}
\left(\left(1-\frac{\vec{p}^2}{8M_{\mathcal{B}_6}^2}\right)\chi,
\frac{\vec{\sigma}\cdot\vec{p}}{2M_{\mathcal{B}_6}}\chi\right)^T,\\
\langle0|\mathcal{B}_{6}^{*\mu}|Qqq(3/2^+)\rangle &=&
\sum_{m_1,m_2}C_{1/2,m_1;1,m_2}^{3/2,m_1+m_2}
\sqrt{2M_{\mathcal{B}_{6}^*}}\nonumber\\&&\times\left(\left(1-\frac{\vec{p}^2}{8M_{\mathcal{B}_6^*}}\right)
\chi_{\frac{1}{2},m_1},
\frac{\vec{\sigma}\cdot\vec{p}}{2M_{\mathcal{B}_6^*}}
\chi_{\frac{1}{2},m_1}\right)^T\epsilon^{\mu}_{m_2}.
\end{eqnarray*}

\subsection{Effective potential}\label{poten}

\begin{figure}[!htbp]
  \centering
  \includegraphics[width=2.5in]{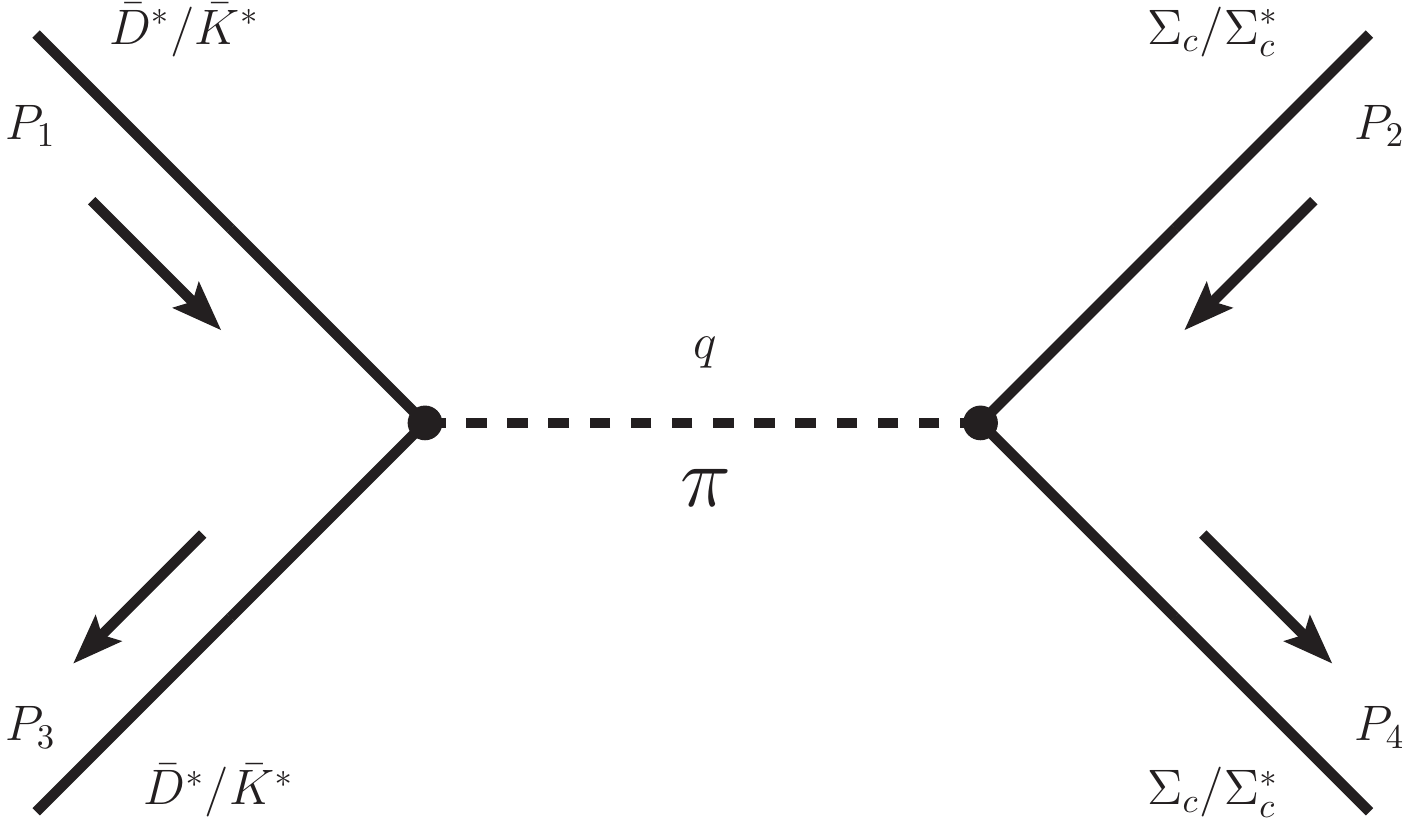}\\
  \caption{The Feynman diagrams for the $\Sigma_c\bar{D}^*$, $\Sigma_c^*\bar{D}^*$, $\Sigma_c\bar{K}^*$ and $\Sigma_c^*\bar{K}^*$ systems in OPE model.}\label{Feyn}
\end{figure}

In the OPE model, we first obtain the scattering amplitude in Fig.
\ref{Feyn}.

The effective potential can be related to the scattering amplitude
with the Breit approximation. For example, the effective potential
for the $\Sigma_c\bar{D}^*$ system in the momentum space is
\begin{eqnarray}\label{breit}
\mathcal{V}_{E}^{\Sigma_c\bar{D}^*\rightarrow
\Sigma_c\bar{D}^*}(\vec{q}) &=&
          -\frac{\mathcal{M}(\Sigma_c\bar{D}^*\rightarrow\Sigma_c\bar{D}^*)}
          {\sqrt{\prod_i2M_i\prod_f2M_f}},
\end{eqnarray}
where $M_i$ and $M_f$ are the masses of the initial and final
states, respectively.
$\mathcal{M}(\Sigma_c\bar{D}^*\rightarrow\Sigma_c\bar{D}^*)$ denotes
the scattering amplitude for the
$\Sigma_c\bar{D}^*\rightarrow\Sigma_c\bar{D}^*$ process by
exchanging one pion meson in t-channel. The effective potential in
the coordinate space $\mathcal{V}(\vec{r})$ is obtained by
performing Fourier transformation to the momentum space effective
potential $\mathcal{V}(\vec{q})$,
\begin{eqnarray}
\mathcal{V}_{E}^{\Sigma_c\bar{D}^*\leftrightarrow
\Sigma_c\bar{D}^*}(\vec{r}) =
          \int\frac{d^3\vec{p}}{(2\pi)^3}e^{i\vec{p}\cdot\vec(r)}
          \mathcal{V}_{E}^{\Sigma_c\bar{D}^*\leftrightarrow \Sigma_c\bar{D}^*}(\vec{q})\mathcal{F}^2(q^2,m_E^2).\nonumber
\end{eqnarray}
In the above expression, the monopole form factor
$\mathcal{F}(q^2,m_E^2)$ is introduced at each interaction vertex to
compensate the off-shell effect of the exchanged pion.
$\mathcal{F}(q^2,m_E^2)=(\Lambda^2-m_E^2)/(\Lambda^2-q^2)$, where
$m_E$ and $q$ are the mass and four-momentum of the exchanged meson,
respectively. $\Lambda$ is the cutoff with the value around one to
several GeV.

The general expressions of the subpotentials for the processes
$\Sigma_c\bar{D}^*\rightarrow \Sigma_c\bar{D}^*$,
$\Sigma_c^*\bar{D}^*\rightarrow \Sigma_c^*\bar{D}^*$,
$\Sigma_c\bar{K}^*\rightarrow \Sigma_c\bar{K}^*$ and
$\Sigma_c^*\bar{K}^*\rightarrow \Sigma_c^*\bar{K}^*$ are
\begin{eqnarray}
\mathcal{V}^{\Sigma_c\bar{D}^*\rightarrow
\Sigma_c\bar{D}^*}_{\pi}(\vec{r}) &=&
            \frac{gg_1}{f_{\pi}^2}\mathcal{V}_1,\label{sub01}\\
\mathcal{V}^{\Sigma_c^*\bar{D}^*\rightarrow
\Sigma_c^*\bar{D}^*}_{\pi}(\vec{r}) &=&
            \frac{3}{2}\frac{gg_1}{f_{\pi}^2}\mathcal{V}_2,\label{sub02}\\
\mathcal{V}^{\Sigma_c\bar{K}^*\rightarrow
\Sigma_c\bar{K}^*}_{\pi}(\vec{r}) &=&
            \frac{g_{\pi\bar{K}^*\bar{K}^*}g_1}{\sqrt{2}f_{\pi}}\mathcal{V}_1,\label{sub03}\\
\mathcal{V}^{\Sigma_c^*\bar{K}^*\rightarrow
\Sigma_c^*\bar{K}^*}_{\pi}(\vec{r}) &=&
            \frac{3}{2\sqrt{2}}\frac{g_{\pi\bar{K}^*\bar{K}^*}g_1}{f_{\pi}}\mathcal{V}_2
            \label{sub04}
\end{eqnarray}
where
\begin{eqnarray}
\mathcal{V}_1 &=&
            \frac{1}{3}\left[\left(i\vec{\epsilon_1}\times\vec{\epsilon_3}^{\dag}\right)
            \cdot\vec{\sigma}\right]\nabla^2Y(\Lambda,m_{\pi},\vec{r})\nonumber\\
            &&+\frac{1}{3}S(\hat{r},i\vec{\epsilon_1}
            \times\vec{\epsilon_3}^{\dag},\vec{\sigma})
            r\frac{\partial}{\partial r}\frac{1}{r}\frac{\partial}{\partial r}Y(\Lambda,m_{\pi},\vec{r}),\label{yy0}\\
\mathcal{V}_2 &=&
            \sum_{a,b,c,d}\left\langle\frac{1}{2},a;1,b\bigg|\frac{3}{2},a+b\right\rangle
            \left\langle\frac{1}{2},c;1,d\bigg|\frac{3}{2},c+d\right\rangle
            \chi^{a\dag}_4\chi^{c}_2\nonumber\\
            &&\times\Big\{\frac{1}{3}
            \left(\vec{\epsilon}_1\times\vec{\epsilon}_3^{\dag}\right)
            \cdot\left(\vec{\epsilon}_2^{d}\times\vec{\epsilon}_4^{b\dag}\right)
            \nabla^2Y(\Lambda,m_{\pi},\vec{r})\nonumber\\
            &&+\frac{1}{3}S\left(\hat{r},\vec{\epsilon}_1\times\vec{\epsilon}_3^{\dag},
            \vec{\epsilon}_2^{d}\times\vec{\epsilon}_4^{b\dag}\right)
            r\frac{\partial}{\partial r}\frac{1}{r}\frac{\partial}{\partial r}Y(\Lambda,m_{\pi},\vec{r})\Big\}.
\end{eqnarray}
In the above expressions, $S(\hat{r},\vec{x},\vec{y})=
3(\hat{r}\cdot\vec{x})(\hat{r}\cdot\vec{y})-\vec{x}\cdot\vec{y}$,
and
\begin{eqnarray}
Y(\Lambda,m,{r}) &=&\frac{1}{4\pi r}(e^{-mr}-e^{-\Lambda
r})-\frac{\Lambda^2-m^2}{8\pi \Lambda}e^{-\Lambda r}.\label{yy}
\end{eqnarray}

To obtain the total effective potentials of the $\Sigma_c\bar{M}^*$
and $\Sigma_c^*\bar{M}^*$ systems, the effective potentials in Eqs.
(\ref{sub01})-(\ref{sub04}) should be sandwiched between the flavor
wave functions. Thus, the general total effective potential for
different systems can be expressed as
\begin{eqnarray}
V_{\text{total}}^{\Sigma_c\bar{M}^*} &=&
          \mathcal{G}\mathcal{V}^{\Sigma_c\bar{M}^*\rightarrow \Sigma_c\bar{M}^*}_{\pi}(\vec{r}),\label{totalv1}\\
V_{\text{total}}^{\Sigma_c^*\bar{M}^*} &=&
           -\mathcal{G}\mathcal{V}^{\Sigma_c^*\bar{M}^*\rightarrow \Sigma_c^*\bar{M}^*}_{\pi}(\vec{r}),\label{totalv2}
\end{eqnarray}
where the isospin factor $\mathcal{G}=1$ for the isospin-1/2 system
and $\mathcal{G}=-1/2$ for the isospin-$3/2$ system. The angular
momentum operators in Eqs. (\ref{sub01})-(\ref{sub04}) will be
replaced by a series of numerical matrixes collected in Table
\ref{matrix}. We need to specify that the S-D mixing effect is considered in our calculation, which makes
the corresponding replacements in Eqs. (\ref{yy0})-(\ref{yy}) just indicated in Table \ref{matrix}.

\renewcommand{\arraystretch}{1.5}
\begin{table}[!htbp]
\centering \caption{The matrix expressions for the angular momentum
operators according to
$\langle{}^{2S'+1}L'_{J'}|\Omega_i|{}^{2S+1}L_{J}\rangle$. Here,
$\Omega_1=(i\vec{\epsilon_1}\times\vec{\epsilon_3}^{\dag})\cdot\vec{\sigma}$,
$\Omega_2=S(\hat{r},i\vec{\epsilon_1}\times\vec{\epsilon_3}^{\dag},\vec{\sigma})$,
$\Omega_3=\sum_{a,b;c,d}C_{1/2,a;1,b}^{3/2,m}C_{1/2,c;1,d}^{3/2,n}\chi^{a\dag}_4\chi^{c}_2
(\vec{\epsilon}_1\times\vec{\epsilon}_3^{\dag})
\cdot(\vec{\epsilon}_2^{d}\times\vec{\epsilon}_2^{b\dag})$ and
$\Omega_4=\sum_{a,b;c,d}C_{1/2,a;1,b}^{3/2,m}C_{1/2,c;1,d}^{3/2,n}\chi^{a\dag}_4\chi^{c}_2
S(\hat{r},\vec{\epsilon}_1\times\vec{\epsilon}_3^{\dag},
\vec{\epsilon}_2^{d}\times\vec{\epsilon}_2^{b\dag})$.}
\label{matrix} \tiny{\begin{tabular}{cccc}
  \toprule[1pt]\toprule[1pt]
  $J$  &$1/2$     &$3/2$   &$5/2$\\
  \midrule[1pt]
  $\Omega_1$    &$\left(\begin{array}{cc}-2 &0\\
                           0 &1\end{array}\right)$
                &$\left(\begin{array}{ccc}1  &0  &0\\
                           0  &-2 &0\\
                           0  &0  &1\end{array}\right)$
           &$\times$\\
  $\Omega_2$    &$\left(\begin{array}{cc}0 &-\sqrt{2}\\
                           -\sqrt{2} &-2\end{array}\right)$
           &$\left(\begin{array}{ccc}0  &1  &2\\
                            1  &0  &-1\\
                            2  &-1 &0\end{array}\right)$
           &$\times$\\
  $\Omega_3$    &$\left(\begin{array}{cc}\frac{5}{3} &0\\
                           0 &\frac{2}{3}\end{array}\right)$
                &$\left(\begin{array}{ccc}\frac{2}{3}  &0  &0\\
                           0  &\frac{5}{3} &0\\
                           0  &0  &\frac{2}{3}\end{array}\right)$
                &$\left(\begin{array}{cccc}-1  &0  &0  &0\\
                           0  &\frac{5}{3} &0  &0\\
                           0  &0  &\frac{2}{3}  &0\\
                           0  &0  &0    &-1\end{array}\right)$\\
  $\Omega_4$     &$\left(\begin{array}{cc}0         &-\frac{7}{3\sqrt{5}}\\
                      -\frac{7}{3\sqrt{5}} &\frac{16}{15}\end{array}\right)$
                 &$\left(\begin{array}{ccc}0  &\frac{7}{3\sqrt{10}}  &-\frac{16}{15}\\
                   \frac{7}{3\sqrt{10}}  &0 &-\frac{7}{3\sqrt{10}}\\
                -\frac{16}{15}  &-\frac{7}{3\sqrt{10}}  &0\end{array}\right)$
                 &\scriptsize{$\left(\begin{array}{cccc}
    0  &\frac{2}{\sqrt{15}}  &\frac{1}{5}\sqrt{\frac{3}{7}}   &-\frac{2\sqrt{14}}{5}\\
\frac{2}{\sqrt{15}}  &0 &\frac{1}{3}\sqrt{\frac{7}{5}}    &-4\sqrt{\frac{2}{105}}\\
\frac{1}{5}\sqrt{\frac{3}{7}}  &\frac{1}{3}\sqrt{\frac{7}{5}}
&-\frac{16}{21}
       &-\frac{1}{7}\sqrt{\frac{2}{3}}\\
-\frac{2\sqrt{14}}{5}   &-4\sqrt{\frac{2}{105}}
&-\frac{1}{7}\sqrt{\frac{2}{3}}
       &-\frac{4}{7}\end{array}\right)$}\\
\bottomrule[1pt] \bottomrule[1pt]
\end{tabular}}
\end{table}

\section{Numerical results}\label{sec3}

With the obtained OPE effective potentials for the
$\Sigma_c\bar{M}^*$ and $\Sigma_c^*\bar{M}^*$ systems with
$\bar{M}^*=(\bar{D}^*, \bar{K}^*)$ listed in Section \ref{sec2}, we
can find the possible bound solutions (the binding energy $E$,
corresponding root-mean-square radius $r_{RMS}$ and radial wave
function $\phi(r)$) by solving the coupled-channel
Schr$\ddot{\text{o}}$dinger equation with the help of the FESSDE
program \cite{Abrashkevichn:1995cj,Abrashkevich:1998cj}. The
corresponding kinetic terms include
\begin{eqnarray}
K_{\Sigma_c\bar{M}^*}^{J=1/2} &=& \text{diag}
                   \left(-\frac{1}{2m_1}\nabla^2, -\frac{1}{2m_1}\nabla_1^2\right),\\
K_{\Sigma_c\bar{M}^*}^{J=3/2} &=& \text{diag}
                   \left(-\frac{1}{2m_{1}}\nabla^2, -\frac{1}{2m_1}\nabla_1^2,
                   -\frac{1}{2m_1}\nabla_1^2\right),\\
K_{\Sigma_c^*\bar{M}^*}^{J=1/2} &=& \text{diag}
                   \left(-\frac{1}{2m_2}\nabla^2, -\frac{1}{2m_2}\nabla_1^2\right),\\
K_{\Sigma_c^*\bar{M}^*}^{J=3/2} &=& \text{diag}
                   \left(-\frac{1}{2m_2}\nabla^2, -\frac{1}{2m_2}\nabla_1^2,
                   -\frac{1}{2m_2}\nabla_1^2\right),\\
K_{\Sigma_c^*\bar{M}^*}^{J=3/2} &=& \text{diag}
                   \Bigg(-\frac{1}{2m_2}\nabla^2, -\frac{1}{2m_2}\nabla_1^2,
                   -\frac{1}{2m_2}\nabla_1^2, -\frac{1}{2m_2}\nabla_1^2\Bigg)
\end{eqnarray}
with $\nabla^2 = \frac{1}{r^2}\frac{\partial}{\partial
r}r^2\frac{\partial}{\partial r}$ and $\nabla_1^2 = \nabla^2-6/r^2$,
where $m_1$ and $m_2$ are the reduced masses of the
$\Sigma_c\bar{M}^*$ and $\Sigma_c^*\bar{M}^*$ system, respectively.

\subsection{$P_c(4380)$ and $P_c(4450)$}

\begin{figure}[htbp]
  \centering
  \includegraphics[width=0.75\textwidth]{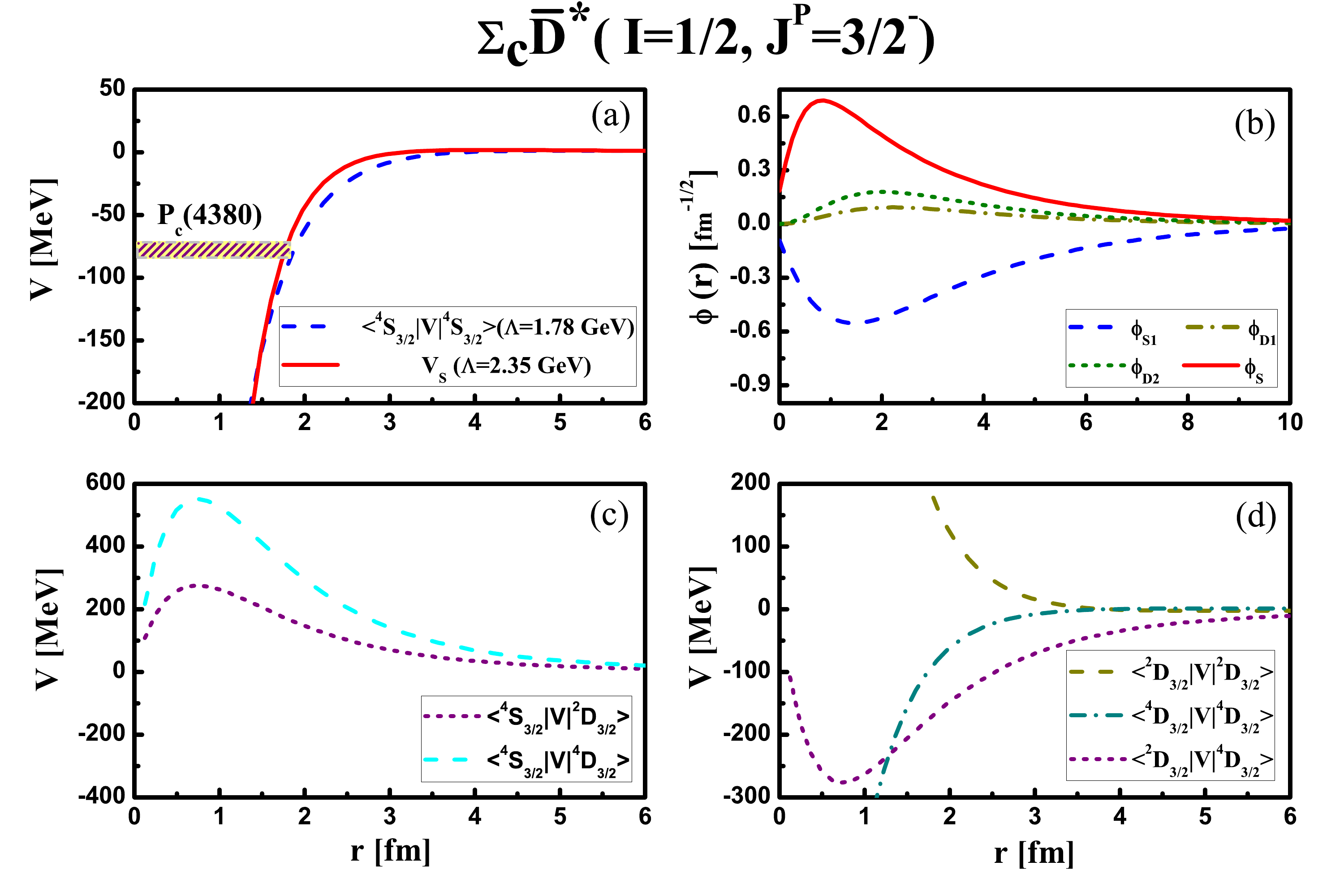}\\
  \includegraphics[width=0.75\textwidth]{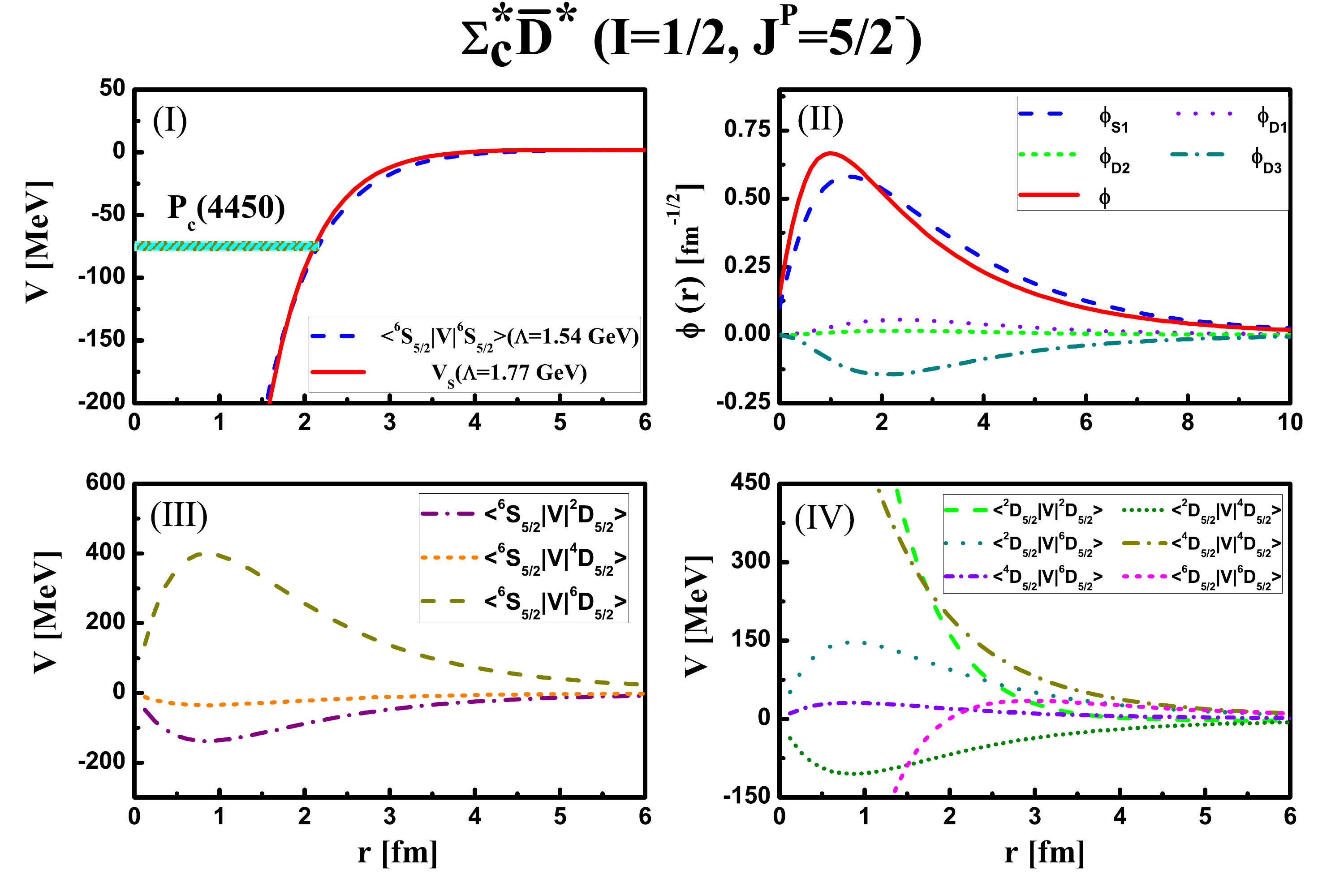}\\
\caption{(color online). The variations of the obtained OPE
effective potentials with $r$ for the $\Sigma_c^{(*)}\bar{D}^*$
systems, and the obtained bound state solutions. Here, $\phi_{S1}$,
$\phi_{D1}$ and $\phi_{D2}$ in Fig. \ref{pc} (b) denote the spatial
wave functions of the $\Sigma_c\bar{D}^*$ ($I=1/2,\,J^P=3/2^-$) state
with the angular wave functions
$|{}^4\mathbb{S}_{\frac{3}{2}}\rangle$,
$|{}^2\mathbb{D}_{\frac{3}{2}}\rangle$ and
$|{}^4\mathbb{D}_{\frac{3}{2}}\rangle$, respectively. The notations
$\phi_S$, $\phi_{D1}$, $\phi_{D2}$ and $\phi_{D3}$ in Fig. \ref{pc}
(II) are the same as in Fig. \ref{pc} (b). }\label{pc}
\end{figure}

In Ref. \cite{Chen:2015loa}, the authors considered the S-wave
contribution only and obtained the bound solutions for the systems
$\Sigma_c\bar{D}^*$ with $(I=1/2, J^P=3/2^-)$ and $\Sigma_c^*\bar{D}^*$
with $(I=1/2, J^P=5/2^-)$, which may correspond to $P_c(4380)$ and
$P_c(4450)$, respectively. In this work, we include the S-D mixing
and restudy these systems. The numerical results for the
$\Sigma_c\bar{D}^*$ system with $(I=1/2, J^P=3/2^-)$ and the
$\Sigma_c^*\bar{D}^*$ system with $(I=1/2, J^P=5/2^-)$ are shown in Fig.
\ref{pc}. For comparison, we also present the results with the
S-wave contribution only in Fig. \ref{pc}.

We compare the total effective potentials with and without the S-D
mixing effect in Fig. \ref{pc} (a) and (I). With the S-D mixing
effect, the spatial wave functions for the $\Sigma_c\bar{D}^*$
system with $(I=1/2, J^P=3/2^-)$ is a $[3\times1]$ column vector with
$\left(|\phi_{S}\rangle, |\phi_{D1}\rangle,
|\phi_{D2}\rangle\right)^T$. As shown in Fig. \ref{pc} (b), the dash
lines with different color stand for the spatial wave functions
$|\phi_{{}^4\mathbb{S}_{\frac{3}{2}}}\rangle$,
$|\phi_{{}^2\mathbb{D}_{\frac{3}{2}}}\rangle$ and
$|\phi_{{}^4\mathbb{D}_{\frac{3}{2}}}\rangle$, which are abbreviated
as $\phi_{S}$, $\phi_{D1}$ and $\phi_{D2}$ in Fig. \ref{pc} (b). The
probability for each component is
$\int|\phi_{i}|^2dr/\sum_{i}\int|\phi_{i}|^2dr$. The S-wave
contribution is dominant and plays a major role in the formation of
molecular pentaquarks.

We notice that the masses of $P_c(4380)$ and $P_c(4450)$ can be
reproduced well under the $\Sigma_c\bar{D}^*$ with $(I=1/2,J^P=3/2^-)$
and $\Sigma_c^*\bar{D}^*$ with $(I=1/2,J^P=5/2^-)$ molecular
assignments, respectively (see Fig. \ref{pc} (a) and (I)). In Fig.
\ref{pc}, the red solid curves stand for the obtained bound state
solutions (effective potentials $V$ [MeV] and radial wave functions
$\phi(r)$ [fm${^{-1/2}}$]) if only considering the S-wave effect. In
this case, $\Lambda=2.35$ GeV and $\Lambda=1.77$ GeV are taken for
the $\Sigma_c\bar{D}^*$ and $\Sigma_c^*\bar{D}^*$ systems,
respectively. The dashed blue curves are the corresponding effective
potentials and radial wave functions for the $\Sigma_c\bar{D}^*$
system with $(I=1/2,J^P=3/2^-)$ and the $\Sigma_c^*\bar{D}^*$ system
with $(I=1/2,J^P=5/2^-)$. Now, the cutoff $\Lambda=1.78$ GeV and
$\Lambda=1.54$ GeV are taken for the $\Sigma_c\bar{D}^*$ system with
$(I=1/2,J^P=3/2^-)$ and the $\Sigma_c^*\bar{D}^*$ system with
$(I=1/2,J^P=5/2^-)$, respectively, in order to reproduce the masses of
$P_c(4380)$ and $P_c(4450)$. After comparing the results with and
without the S-D mixing effect, we find that the cutoff values become
smaller with the S-D mixing effect. In other words, the S-D mixing
effect is indeed helpful to the formation of these bound states.

We also notice that there exists slight difference of the fitted
$\Lambda$ values for the $\Sigma_c\bar{D}^*$ and
$\Sigma_c^*\bar{D}^*$, which are used to reproduce the central
values of masses of $P_c(4380)$ and $P_c(4450)$. If considering the
experimental errors for their mass measurement, the difference of
the fitted $\Lambda$ values for the $\Sigma_c\bar{D}^*$ and
$\Sigma_c^*\bar{D}^*$ becomes subtle.

Under the S-wave hidden-charm molecular state assignment to
$P_c(4380)$ and $P_c(4450)$, the parities of $P_c(4380)$ and
$P_c(4450)$ are negative. However,  the LHCb's measurement suggests
that $P_c(4380)$ and $P_c(4450)$ have opposite parities
\cite{Aaij:2015tga}. Facing such situation, we need to clarify this
point. Under the molecular state scheme, there may also exist the
P-wave, D-wave or even higher orbital excitations if the binding
energy of the lowest S-wave hadronic molecule reaches up to several
tens of MeV. The P-wave state has an excitation energy around
several to tens of MeV, which is slightly higher than that of the
S-wave ground state. Considering this status, there exists the
possibility that the S-wave and P-wave states may completely overlap
with each other, i.e., two or more resonant signals around 4450 MeV
may exist, where these two states are close to each other but may
carry different parity. If the P-wave or higher excitation is very
broad, such a state may easily be mistaken as the background.

While the observed $P_c(4380)$ corresponds to the discussed S-wave
$\Sigma_c\bar{D}^*$ molecular state, the observed $P_c(4450)$ may be
a P-wave excitation. There may also exist a very broad S-wave
$\Sigma_c^*\bar{D}^*$ molecular state around 4450 MeV which was
discussed in the present work. This broad S-wave state around 4450
MeV cannot be distinguished from the background.  The fact that the
different assignment of the spin and parity of these two $P_c$
states yields roughly the same good fit \cite{Chen:2015loa} can be
reasonably understood.

On the other hand, if the P-wave excitation lies several MeV within
4380 MeV and has a width as narrow as several MeV, it will be hard
to identify this state since it may probably be buried by the broad
$P_c(4380)$ resonance with 205 MeV width. Thus, we also suggest
future experiment to collect a huge amount of experimental data to
identify the nearly degenerate resonances with different parities
and widths.

\subsection{Other hidden-charm molecular pentaquarks}

Besides $P_c(4380)$ and $P_c(4450)$, we further predict other
possible hidden-charm $\Sigma_c^{(*)}\bar{D}^*$ molecular
pentaquarks and collect the corresponding results in Table
\ref{numpc}. Our numerical results are rather sensitive to the
cutoff parameter $\Lambda$. In this work, we present the bound state
solutions by scanning a $\Lambda<5$ GeV range from the experience of
the nuclear force \cite{Yukawa:1935xg,Machleidt:1989tm,Machleidt:1987hj,Epelbaum:2008ga}.

\renewcommand{\arraystretch}{1.5}
\begin{table}[!htbp]
  \centering
\caption{The typical values of the obtained bound state solutions
$[\Lambda, E, r_{RMS}]$ for the hidden-charm $\Sigma_c\bar{D}^*$ and
$\Sigma_c^*\bar{D}^*$ systems with all possible configurations and $(I,J^P)$ quantum numbers.
Here, $E$, $r_{RMS}$, and $\Lambda$ are in units of MeV, fm, and
GeV, respectively. The masses of these discussed molecular states can be determined by relation $M_{the}+E$. Here, threshold energy $M_{the}$ is taken as 4462 MeV and 4527 MeV for the $\Sigma_c \bar{D}^*$ and $\Sigma_c^* \bar{D}^*$ systems, respectively.}\label{numpc}
  \scriptsize{\begin{tabular}{cc|ccc}
     \toprule[1pt]\toprule[1pt]
     \multicolumn{2}{c}{$\Sigma_c\bar{D}^*$}    &\multicolumn{3}{|c}{$\Sigma_c^*\bar{D}^*$}\\\midrule[1pt]
  $(1/2,1/2^-)$   &   &$(1/2,1/2^-)$   &$(1/2,3/2^-)$  &
  \\\midrule[0.6pt]
   $[2.53, -2.42, 2.26]$   &
     &$[2.72, -2.43, 2.31]$   &$[2.18, -2.48, 2.26]$    &\\
                              $[2.65, -7.58, 1.38]$    &
     &$[2.92, -8.73, 1.35]$   &$[2.38, -9.10, 1.33]$    &\\\midrule[1pt]
  $(I,J^P)=(3/2,1/2^-)$   &$(3/2,3/2^-)$   &$(3/2,1/2^-)$   &$(3/2,3/2^-)$  &$(3/2,5/2^-)$\\
  \midrule[0.6pt]
  $[1.73,-2.55,1.95]$   &$[4.16, -2.27, 2.20]$
     &$[1.44,-2.65,1.92]$  &$[2.36, -1.80, 2.33]$  & $[3.76, -2.54, 2.11]$\\
  $[1.85,-8.91,1.10]$ &$[4.38, -7.06, 1.32]$
  &$[1.56, -9.81,1.07]$  &$[2.56, -9.78, 1.08]$     & $[3.96, -7.48, 1.32]$\\
  \bottomrule[1pt]\bottomrule[1pt]
  \end{tabular}}
\end{table}

If the cutoff is allowed to vary in the range $0.5\sim 5$ GeV, there
exist bound solutions for the $\Sigma_c\bar{D}^*$ systems with
$(I,J^P)=$$(1/2,1/2^-)$, $(1/2,3/2^-)$, $(3/2,1/2^-)$ and $(3/2,3/2^-)$, and
the $\Sigma_c^*\bar{D}^*$ systems with $(I,J^P)=$ $(1/2,1/2^-)$,
$(1/2,3/2^-)$, $(1/2,5/2^-)$, $(3/2,1/2^-)$, $(3/2,3/2^-)$ and $(3/2,5/2^-)$.

If the cutoff is fixed as $\Lambda=1.78$ GeV in order to reproduce
the mass of $P_c(4380)$ under the $\Sigma_c\bar{D}^*$ $(I=1/2,
J^P=3/2^-)$ assignment, there exists a shallow $\Sigma_c\bar{D}^*$
$(I=3/2, J^P=1/2^-)$ molecular state, which is marked as $P'_c(4460)$ in
Table \ref{numpc}, where the obtained energy is $E\simeq -5$ MeV.
For the $\Sigma_c^*\bar{D}^*$ system, we find the bound state
solution ($E\simeq -10$ MeV) for the $\Sigma_c^*\bar{D}^*$ system
with $(I=3/2, J^P=1/2^-)$ with the cutoff $\Lambda=1.54$ GeV which is
adopted to reproduce the mass of $P_c(4450)$. This state is named as
$P'_c(4520)$.

With $\Lambda=1.78$ GeV for the $\Sigma_c\bar{D}^*$ systems and
$\Lambda=1.54$ GeV for the $\Sigma_c^*\bar{D}^*$ systems, we cannot
find a bound state solution for the $\Sigma_c\bar{D}^*$ systems
with $(I,J^P)=(1/2,1/2^-)$ and $(3/2,3/2^-)$, the $\Sigma_c^*\bar{D}^*$
systems with $(I,J^P)=(1/2,1/2^-), (1/2,3/2^-),(3/2,3/2^-)$ and $(3/2,5/2^-)$.
This observation is almost the same as in Ref. \cite{Chen:2015loa}
where only the S-wave is taken into account .

In short summary, there exist at least two hidden-charm shallow
molecular pentaquarks $P'_c(4460)$ and $P'_c(4520)$, which are the
partners of $P_c(4380)$ and $P_c(4450)$. In Table \ref{decaypc}, we
list these allowed decay modes of $P_c(4380)$, $P_c(4450)$,
$P_c'(4460)$ and $P_c'(4520)$.

\renewcommand{\arraystretch}{1.5}
\begin{table}[!htbp]
\centering \caption{The allowed decay channels for $P_c(4380)$,
$P_c(4450)$, $P_c'(4460)$ and $P_c'(4520)$. Here, $S$ and $D$ stand
for the S-wave and D-wave decay mode, respectively. And $\times$
indicates that this decay process is forbidden.}\label{decaypc}
\scriptsize{\begin{tabular}{ccccc} \toprule[1pt] Decay channels
&$\Sigma_c\bar{D}^*(1/2, 3/2^-)$   &$\Sigma_c\bar{D}^*(3/2, 1/2^-)$   &$\Sigma_c^*\bar{D}^*(1/2, 5/2^-)$   &$\Sigma_c^*\bar{D}^*(3/2 ,1/2^-)$\\
\midrule[1pt]
$\Lambda_c\bar{D}$   &$D$    &$\times$     &$D$    &$\times$\\
$\Lambda_c\bar{D}^*$  &$S/D$    &$\times$     &$D$    &$\times$\\
$\Sigma_c\bar{D}$    &$D$    &$\times$     &$D$    &$\times$\\
$\eta_c N$           &$D$    &$\times$     &$D$    &$\times$\\
$J/\psi N$           &$S$    &$\times$     &$D$    &$\times$\\
$\eta_c \Delta$      &$D$    &$S$  &$D$    &$S$\\
$J/\psi \Delta$      &$S$    &$S$  &$D$    &$S$\\
\bottomrule[1pt]
\end{tabular}}
\end{table}

\subsection{Charm-strange molecular pentaquarks}

The numerical results of the $\Sigma_c\bar{K}^*$ and
$\Sigma_c^*\bar{K}^*$ systems with all possible quantum numbers are
listed in Table \ref{numpcs}, where the cutoff is allowed to vary in
the range $\Lambda=0.5 \sim5$ GeV. There exist bound state solutions
for the $\Sigma_c\bar{K}^*$ systems with $(I=1/2, J^P=1/2^-)$ and
$(I=3/2,J^P=1/2^-)$, and the $\Sigma_c^*\bar{K}^*$ systems with $(I=1/2,
J^P=1/2^-)$, $(I=1/2, J^P=3/2^-)$, $(I=3/2, J^P=1/2^-)$ and $(I=3/2, J^P=3/2^-)$.

\renewcommand{\arraystretch}{1.5}
\begin{table}[!htbp]
  \centering
\caption{The typical values of the obtained bound state solutions
$[\Lambda, E, r_{RMS}]$ for the $\Sigma_c\bar{K}^*$ and
$\Sigma_c^*\bar{K}^*$ systems. Here, $E$, $r_{RMS}$, and $\Lambda$
are in units of MeV, fm, and GeV, respectively. And $[\times]$
indicates that there does not exist a bound state
solution. The masses of these discussed molecular states can be determined by relation $M_{the}+E$. Here, threshold energy $M_{the}$ is taken as 3345 MeV and 3410 MeV for the $\Sigma_c \bar{K}^*$ and $\Sigma_c^* \bar{K}^*$ systems, respectively.}\label{numpcs}
  \scriptsize{\begin{tabular}{cc|ccc}
     \toprule[1pt]\toprule[1pt]
     \multicolumn{2}{c}{$\Sigma_c\bar{K}^*$}    &\multicolumn{3}{|c}{$\Sigma_c^*\bar{K}^*$}\\\midrule[1pt]
  $(I,J^P)=(1/2,1/2^-)$   &$(1/2,3/2^-)$   &$(1/2,1/2^-)$   &$(1/2,3/2^-)$  &$(1/2,5/2^-)$
  \\\midrule[0.6pt]
  $[3.84, -2.24, 2.83]$       &$[1.78,-3.88,2.16]$
  &$[4.14, -2.13, 2.94]$   &$[3.22, -2.11, 2.91]$   &$[1.05,-2.43,2.16]$\\
  $[3.98, -9.58, 1.46]$     &$[1.84, -7.31,1.63]$
  &$[4.34, -9.18, 1.52]$   &$[3.42, -9.24, 1.52]$   &$[1.17,-9.66,1.22]$
  \\\midrule[1pt]
  $(I,J^P)=(3/2,1/2^-)$   &$(3/2,3/2^-)$   &$(3/2,1/2^-)$   &$(3/2,3/2^-)$  &$(3/2,5/2^-)$\\
  \midrule[0.6pt]
  $[2.52, -2.20, 2.63]$   & $[\times]$
  &$[2.08, -2.63, 2.43]$   &$[3.56, -2.49, 2.51]$    & $[\times]$\\
  $[2.62, -7.77, 1.45]$   & $[\times]$
  &$[2.18, -9.09, 1.36]$   & $[3.70, -9.04, 1.37]$    & $[\times]$\\
  \bottomrule[1pt]\bottomrule[1pt]
  \end{tabular}}
\end{table}

If we take $\Lambda=1.77$ GeV for the $\Sigma_c\bar{K}^*$ systems
and $\Lambda=$1.54 GeV for the $\Sigma_c^*\bar{K}^*$ systems from
the experience of $P_c(4380)$ and $P_c(4450)$, there only exist two
charm-strange molecular pentaquarks $P_{cs}(3340)$ and
$P_{cs}(3400)$ corresponding to the $\Sigma_c\bar{K}^*$ $(I=1/2,
J^P=3/2^-)$ molecular pentaquark with $E_{bin}=-3.88$ MeV and the
$\Sigma_c^*\bar{K}^*$ $(I=1/2, J^P=5/2^-)$ system with
$E_{bin}=-6.94$ MeV, respectively. We denote these two charm-strange
molecular pentaquarks as $P_{cs}(3340)$ and $P_{cs}(3400)$, where
the numbers 3340 MeV and 3400 MeV are their mass deduced from the
relations $m_{\Sigma_c}+m_{\bar{K}^*}+E_{bin}$ and
$m_{\Sigma_c^*}+m_{\bar{K}^*}+E_{bin}$, respectively. In other
words, we adopt their masses to name these two predicted
charm-strange molecular pentaquarks. The possible decay channels of
$P_{cs}(3340)(I=1/2, J^P=3/2^-)$ and
$P_{cs}(3400)(I=1/2, J^P=5/2^-)$ are shown in Table
\ref{decaypcs}. As shown in Fig. \ref{pcs} (b), $P_{cs}(3340)$ and
$P_{cs}(3400)$ can be produced via the $\Lambda_b\rightarrow
\bar{D}^0P_{cs}^0$ decay at LHCb.

\renewcommand{\arraystretch}{1.5}
\begin{table}[!htbp]
\centering \caption{The allowed decay channels of
$P_{cs}(3340)(I=1/2, J^P=3/2^-)$ and
$P_{cs}(3400)(I=1/2, J^P=5/2^-)$. The $S$, $P$, $D$ and
$F$ denote the S-wave, P-wave, D-wave and F-wave decay modes
respectively.}\label{decaypcs}
\scriptsize{\begin{tabular}{ccccccccccc} \toprule[1pt]\toprule[1pt]
 &$D\Lambda$    &$D\Lambda(1405)$    &$D\Lambda(1520)$    &$D\Sigma$   &$D\Sigma(1385)$    &$D\Sigma(1660)$  &$D^*\Lambda/\Sigma$
\\\midrule[0.6pt]
$P_{cs}(3340)$    &$D$      &$P$     &$P$   &$D$   &$S$   &$D$    &$S$\\
$P_{cs}(3400)$    &$D$      &$F$     &$P$   &$D$   &$D$   &$D$
&$D$
\\\midrule[1pt]
    &$\bar{K}\Lambda_c$    &$\bar{K}\Lambda_c(2595)$   &$\bar{K}\Lambda_c(2625)$     &$\bar{K}\Sigma_{c}$   &$\bar{K}\Sigma_c^*$
\\\midrule[0.6pt]
$P_{cs}(3340)$      &$D$    &$P$    &$P$    &$D$    &$S$\\
$P_{cs}(3400)$      &$D$    &$F$    &$P$    &$D$    &$D$
\\\midrule[1pt]
   &$\Xi_c\pi/\eta$    &$\Xi_c\rho/\omega$       &$\Xi_c'\pi/\eta$    &&$\Xi_c(2645)\pi/\eta$      &$\Xi_c(2790)\pi$ &$\Xi_c(2815)\pi$\\\midrule[0.6pt]
$P_{cs}(3340)$    &$D$   &$S$   &$D$     &   &$S$   &$P$   &$P$\\
$P_{cs}(3400)$    &$D$   &$D$   &$D$     &   &$D$   &$F$   &$P$\\
\bottomrule[1pt]\bottomrule[1pt]
\end{tabular}}
\end{table}

In the search of the charm-strange molecular pentaquarks
$P_{cs}(3340)$ and $P_{cs}(3400)$, it is important to distinguish
$P_{cs}(3340)$ and $P_{cs}(3400)$ from the excited P-wave
charm-strange baryons $\Xi_c(1P, 2P)$ etc. The charm-strange baryon
$\Xi_c(3123)$ was observed by the BABAR Collaboration
\cite{Aubert:2007dt}. Its mass is around 220-280 MeV lower than
$P_{cs}(3340)$ and $P_{cs}(3400)$.

In the relativistic quark-diquark picture, Ebert {\it et al.}
predicted the masses of the charm-strange baryons
\cite{Ebert:2007nw}. The mass of $\Xi_c(2P)$ with the scalar diquark
$S_{qq}=0$ and $(I=1/2, J^P=3/2^-)$ is 3199 MeV. For
the $(I=1/2, J^P=5/2^-)$ $\Xi_c$ state with the
vector diquark $S_{qq}=1$, the 1P and 2P masses are 2921 MeV and
3282 MeV for $2P$ state respectively. Fortunately, the above two
charmed baryons predicted in Ref. \cite{Ebert:2007nw} do not overlap
with $P_{cs}(3340)$ and $P_{cs}(3400)$.

However, the mass of the $(I=1/2, J^P=5/2^-)$ $\Xi_c(1F)$
charm-strange baryon with $S_{qq}=1$ was predicted to be around 3.4
GeV \cite{Ebert:2007nw}, which overlaps with the $P_{cs}(3400)$. In
this case, the experimental identification of the $P_{cs}(3400)$
signal will be very challenging. The overpopulation of the the
$(I=1/2, J^P=5/2^-)$ charm-strange baryon around 3400 MeV
may provide some hints. On the other hand, the 1F charm-strange
baryon and $P_{cs}(3400)$ may have very different decay patterns
because of their different internal structures. Further experimental
and theoretical efforts will be helpful to search for the
charm-strange molecular pentaquarks.

\subsection{Hidden-bottom and $B_c$-like molecular pentaquarks}

In Ref. \cite{Wu:2010rv}, Wu and Zou once predicted the
hidden-bottom pentaquark states. where they adopted the meson-baryon
coupled channel unitary approach with the local hidden gauge
formalism.

In the present work, we extend the obtained OPE effective potentials
to study the hidden-bottom
$\Sigma_b^{(*)}B^*$ and $B_c$-like $\Sigma_b^{(*)}\bar{D}^*$ and
$\Sigma_c^{(*)}B^*$ molecular pentaquarks. The reduced masses of the
hidden-bottom and $B_c$-like molecular systems are heavier than
those of the hidden-charm molecular system. Hence the kinetic
energies of the hidden-bottom and $B_c$-like molecular pentaquark
systems are significantly smaller than those of the hidden-charm
molecular pentaquark system. Therefore, the hidden-bottom and
$B_c$-like molecular systems are bound more tightly than the
hidden-charm molecular systems. We collect the bound state solutions
of the hidden-bottom and $B_c$-like molecular pentaquark systems in
Table \ref{numbc}.

\renewcommand{\arraystretch}{1.5}
\begin{table}[!htbp]
  \centering
\caption{The obtained bound state solutions $[\Lambda, E, r_{RMS}]$
for the hidden-bottom and $B_c$-like molecular pentaquark systems.
Here, $E$, $r_{RMS}$, and $\Lambda$ are in units of MeV, fm, and
GeV, respectively. The masses of these discussed molecular states can be determined by relation $M_{the}+E$. Here, threshold energy $M_{the}$ is taken as 11139 MeV, 7779 MeV, 7822 MeV, 11159 MeV, 7843 MeV and 7842 MeV for the $\Sigma_b {B}^*$, $\Sigma_c {B}^*$, $\Sigma_b \bar{D}^*$, $\Sigma_b^* {B}^*$, $\Sigma_c^* {B}^*$, $\Sigma_b^* \bar{D}^*$ systems, respectively.}\label{numbc}
  \scriptsize{\begin{tabular}{cccc}
     \toprule[1pt]\toprule[1pt]
     $(I, J^P)$   &{$\Sigma_b{B}^*$}   &{$\Sigma_c{B}^*$} &{$\Sigma_b\bar{D}^*$}
     \\\midrule[0.6pt]
      $(1/2, 1/2^-)$
                     & $[1.21, -2.39, 1.84]$      & $[1.78, -2.67, 1.93]$
                     & $[1.95, -2.38, 2.09]$ \\
                     & $[1.36, -9.09, 1.14]$        & $[1.93, -9.69, 1.17]$    & $[2.10, -9.07, 1.22]$ \\
      $(1/2, 3/2^-)$
                     & $[0.72, -2.63, 1.65]$      & $[0.94, -2.39, 1.94]$    & $[1.02, -2.68, 1.92]$ \\
                     & $[0.84, -8.91, 1.05]$   & $[1.06, -8.32, 1.19]$    & $[1.14, -8.92, 1.19]$ \\
      $(3/2, 1/2^-)$
                     & $[0.92, -2.72, 1.35]$          & $[1.25, -2.08, 1.81]$    & $[1.37, -2.36, 1.79]$ \\
                     & $[1.04, -8.72, 0.84]$      & $[1.40, -9.58, 0.93]$          & $[1.49, -8.21, 1.03]$ \\
      $(3/2, 3/2^-)$
                     & $[2.01, -2.81, 1.54]$
                     & $[2.90, -2.22, 1.93]$      & $[3.20, -2.22, 2.00]$ \\
                     & $[2.31, -9.89, 0.95]$
                     & $[3.20, -8.83, 1.08]$      & $[3.50, -8.99, 1.10]$ \\
     \midrule[1pt]
     $(I, J^P)$      &{$\Sigma_b^*{B}^*$}   &{$\Sigma_c^*{B}^*$} &{$\Sigma_b^*\bar{D}^*$}\\\hline
      $(1/2, 1/2^-)$
                     & $[1.32, -2.40, 1.92]$         & $[1.90, -2.47, 2.06]$
                     & $[2.11, -2.27, 2.20]$ \\
                     & $[1.53, -8.80, 1.22]$         & $[2.10, -8.49, 1.28]$
                     & $[2.35, -9.88, 1.22]$ \\
      $(1/2, 3/2^-)$
                     & $[1.15, -2.80, 1.80]$         & $[1.55, -2.11, 2.17]$
                     & $[1.37, -2.51, 2.16]$ \\
                     & $[1.35, -9.55, 1.18]$         & $[1.75, -2.00, 1.30]$
                     & $[1.58, -9.63, 1.32]$ \\
      $(1/2, 5/2^-)$
                     & $[0.62, -2.13, 1.74]$        & $[0.80, -2.17, 1.98]$
                     & $[0.86, -1.95, 2.15]$ \\
                     & $[0.74, -9.12, 1.00]$      & $[0.92, -9.04, 1.12]$
                     & $[0.98, -8.54, 1.19]$ \\
      $(3/2, 1/2^-)$
                     & $[0.79, -2.43, 1.44]$        & $[1.05, -2.07, 1.82]$
                     & $[1.15, -2.02, 1.94]$ \\
                     & $[0.91, -8.98, 0.85]$
                     & $[1.15, -6.89, 1.08]$        & $[1.27, -8.23, 1.05]$ \\
      $(3/2, 3/2^-)$
                     & $[1.20, -2.10, 1.56]$        & $[1.70, -2.35, 1.75]$
                     & $[1.90, -2.75, 1.71]$ \\
                     & $[1.40, -9.65, 0.84]$        & $[1.85, -7.79, 1.04]$
                    & $[2.05, -8.70, 1.04]$ \\
      $(3/2, 5/2^-)$
                     & $[1.80, -2.44, 1.69]$        & $[2.59, -2.12, 2.00]$
                     & $[2.23, -2.06, 2.15]$ \\
                     & $[2.10, -9.98, 1.00]$        & $[2.89, -9.46, 1.08]$
                     & $[2.53, -9.68, 1.15]$
     \\\bottomrule[1pt]\bottomrule[1pt]
   \end{tabular}}
\end{table}

For the $\Sigma_b B^*$, $\Sigma_c B^*$, $\Sigma_b \bar{D}^*$ and
$\Sigma_c\bar{D}^*$ systems with the same $(I,J^P)$ quantum number,
the corresponding cutoff values satisfy the relation
$\Lambda_{\Sigma_bB^*}<\Lambda_{\Sigma_cB^*}<\Lambda_{\Sigma_b\bar{D}^*}
<\Lambda_{\Sigma_c\bar{D}^*}$ if we require the same binding energy.
Similarly, we have
$\Lambda_{\Sigma_b^*B^*}<\Lambda_{\Sigma_c^*B^*}<\Lambda_{\Sigma_b^*\bar{D}^*}
<\Lambda_{\Sigma_c^*\bar{D}^*}$ for the $\Sigma_b^* B^*$,
$\Sigma_c^* B^*$, $\Sigma_b^* \bar{D}^*$ and $\Sigma_c^*\bar{D}^*$
systems with the same $(I,J^P)$ quantum number and the same binding
energy. In Table \ref{decaybc}, we list the allowed decay modes of
these hidden-bottom and $B_c$-like molecular pentaquarks.

\renewcommand{\arraystretch}{1.5}
\begin{table}[!htbp]
\centering \caption{The allowed decay channels for the hidden-bottom
$\Sigma_b^{(*)}B^*$ and $B_c$-like
$\Sigma_b^{(*)}\bar{D}^*/\Sigma_c^{(*)}B^*$ molecular
pentaquarks.}\label{decaybc}
\scriptsize{\begin{tabular}{c|ccl} \toprule[1pt]\toprule[1pt]
Systems   &$(I, J^P)$    &Waves &\multicolumn{1}{c}{Decay
channels}\\\midrule[1pt]
\multirow{5}*{$\Sigma_bB^*$}
   &$(1/2, 1/2^-)$
       &$S$    &$\Lambda_bB$, $\Lambda_bB^*$, $\Sigma_bB$, $\Upsilon(1S)N$, $\Upsilon(1S)N(1440)$, $\Upsilon(2S)N$\\
     & &$P$    &$\chi_{b0}(1P)N$, $\chi_{b1}(1P)N$, $h_b(1P)N$, $\chi_{b2}(1P)N$\\
       &$(1/2, 3/2^-)$
       &$S$    &$\Lambda_bB^*$, $\Upsilon(1S)N$, $\Upsilon(1S)N(1440)$, $\Upsilon(2S)N$\\
     & &$P$    &$\chi_{b0}(1P)N$, $\chi_{b1}(1P)N$, $h_b(1P)N$, $\chi_{b2}(1P)N$\\
      & &$D$   &$\Lambda_bB$, $\Sigma_bB$\\
     &$(3/2, 1/2^-|3/2^-)$
       &$S$    &$\Upsilon(1S)\Delta$\\\midrule[1pt]
\multirow{9}*{$\Sigma_b^*B^*$}    &$(1/2, 1/2^-)$
         &$S$   &$\Lambda_bB$, $\Lambda_bB^*$, $\Sigma_bB$, $\Sigma_bB^*$, \\ &&&$\Upsilon(1S)N$, $\Upsilon(1S)N(1440)$, $\Upsilon(2S)N$\\
         &&$P$  &$\chi_{b0}(1P)N$, $\chi_{b1}(1P)N$, $h_b(1P)N$\\
         &&$D$  &$\Sigma_b^*B$, $\Upsilon(1D)N$\\
     &$(1/2, 3/2^-)$
          &$S$  &$\Lambda_bB^*$, $\Sigma_bB^*$, $\Sigma_b^*B$,\\
           &&&$\Upsilon(1S)N$, $\Upsilon(1S)N(1440)$, $\Upsilon(2S)N$, $\Upsilon(1D)N$\\
         &&$P$  &$\chi_{b0}(1P)N$, $\chi_{b1}(1P)N$, $h_b(1P)N$\\
         &&$D$   &$\Lambda_bB$, $\Sigma_bB$\\
     &$(1/2, 5/2^-)$
          &$S$   &$\Upsilon(1D)N$\\
         &&$P$   &$\chi_{b1}(1P)N$, $h_b(1P)N$\\
         &&$D$   &$\Lambda_bB$, $\Lambda_bB^*$, $\Sigma_bB$, $\Sigma_bB^*$, $\Sigma_b^*B$,\\
         &&& $\Upsilon(1S)N$, $\Upsilon(1S)N(1440)$, $\Upsilon(2S)N$\\
     &$(3/2,1/2^-|3/2^-|5/2^-)$
          &$S$   &$\Upsilon(1S)\Delta$\\\midrule[1pt]
\multirow{4}*{$\Sigma_cB^*|\Sigma_b\bar{D}^*$}
     &$(1/2, 1/2^-)$
          &$S$  &$\Lambda_cB|\Lambda_b\bar{D}$, $\Lambda_cB^*|\Lambda_b\bar{D}^*$, $\Sigma_cB|\Sigma_b\bar{D}$,\\
           &&&$B_c|\bar{B}_c(0^-)N$, $B_c|\bar{B}_c(1^-)N$\\
     &$(1/2, 3/2^-)$
          &$S$    &$\Lambda_cB^*|\Lambda_b\bar{D}^*$, $B_c|\bar{B}_c(1^-)N$\\
          &&$D$   &$\Lambda_cB|\Lambda_b\bar{D}$, $B_c|\bar{B}_c(0^-)N$\\
     &$(3/2, 1/2^-|3/2^-)$
          &$S$    &$B_c|\bar{B}_c(1^-)\Delta$\\\midrule[1pt]
\multirow{8}*{$\Sigma_c^*B^*|\Sigma_b^*\bar{D}^*$}
     &$(1/2, 1/2^-)$
          &$S$    &$\Lambda_cB|\Lambda_b\bar{D}$, $\Lambda_cB^*|\Lambda_b\bar{D}^*$, $\Sigma_cB|\Sigma_b\bar{D}$, $\Sigma_cB^*|\Sigma_b\bar{D}^*$, \\ &&& $\Sigma_c^*B|\Sigma_b^*\bar{D}$,
$B_c|\bar{B}_c(0^-)N$, $B_c|\bar{B}_c(1^-)N$\\
     &$(1/2, 3/2^-)$
          &$S$    &$\Lambda_cB^*|\Lambda_b\bar{D}^*$, $\Sigma_cB^*|\Sigma_b\bar{D}^*$, $\Sigma_c^*B|\Sigma_b^*\bar{D}$, $B_c|\bar{B}_c(1^-)N$\\
          &&$D$   &$\Lambda_cB|\Lambda_b\bar{D}$, $\Sigma_cB|\Sigma_b\bar{D}$, $B_c|\bar{B}_c(0^-)N$\\
     &$(1/2, 5/2^-)$
           &$D$   &$\Lambda_cB|\Lambda_b\bar{D}$, $\Lambda_cB^*|\Lambda_b\bar{D}^*$, $\Sigma_cB|\Sigma_b\bar{D}$, $\Sigma_cB^*|\Sigma_b\bar{D}^*$, \\ &&& $\Sigma_c^*B|\Sigma_b^*\bar{D}$,
$B_c|\bar{B}_c(0^-)N$, $B_c|\bar{B}_c(1^-)N$\\
     &$(3/2, 1/2^-)$
          &$S$    &$B_c|\bar{B}_c(1^-)\Delta$\\
     &$(3/2, 3/2^-)$
          &$S$    &$B_c|\bar{B}_c(0^-)\Delta$, $B_c|\bar{B}_c(1^-)\Delta$\\
     &$(3/2, 5/2^-)$
           &$D$   &$B_c|\bar{B}_c(0^-)\Delta$, $B_c|\bar{B}_c(1^-)\Delta$\\
\bottomrule[1pt]\bottomrule[1pt]
\end{tabular}}
\end{table}

We need to emphasize that several states with the same quantum
numbers and very small binding energy appear (see Table
\ref{numbc}). If assuming that these states would have a width
similar to that of the two observed $P_c$ states so far
experimentally, the mass gap between these bound states would be
much smaller than their widths. Thus, it might be hard to identify
them experimentally.

\section{Summary}\label{sec4}

Inspired by the observation of $P_c(4380)$ and $P_c(4450)$ by LHCb
\cite{Aaij:2015tga}, we have investigated the possible molecular
pentaquarks composed of a charmed baryon ($\Sigma_c,\,\Sigma_c^*$)
and a $\bar{D}^*$ meson in the framework of the OPE model, where the
S-D mixing effect is included in our calculation. Our result
indicates the $\Sigma_c\bar{D}^*$ molecular state with $(I=1/2,
J^P=3/2^-)$ and the $\Sigma_c^*\bar{D}^*$ molecular state with
$(I=1/2, J^P=5/2^-)$ have the same mass range as that of the
observed $P_c(4380)$ and $P_c(4450)$, respectively. We also predict
their two partner states composed of the $\Sigma_c\bar{D}^*$ with
$(I=3/2, J^P=1/2^-)$ and $\Sigma_c^*\bar{D}^*$ with $(I=3/2,
J^P=1/2^-)$. We extend the same formalism and predict the
hidden-bottom/$B_c$-like molecular pentaquarks and discuss their
strong decay modes.

As a byproduct, we have also studied the charm-strange molecular
pentaquarks composed of a charmed baryon ($\Sigma_c,\,\Sigma_c^*$)
and a $\bar K^*$ meson. We predict two charmed-strange molecular
pentaquarks $P_{cs}(3340)$ and $P_{cs}(3400)$, which have the
configurations $\Sigma_c\bar{K}^*$ with
$(I=1/2, J^P=3/2^-)$ and $\Sigma_c^*\bar{K}^*$ with
$(I=1/2, J^P=5/2^-)$, respectively. These states can
be produced through the $\Lambda_b\to P_{cs}^0 \bar D^0$ decay
process in Fig. \ref{pcs} (b). The production rate is of the same
order as that of $P_c(4380)$ and $P_c(4450)$. Hopefully these states
can be searched for at LHCb.

In our calculation, we adopt the OPE model, where only one pion
exchange is considered. In fact, for these discussed systems, the
light vector meson ($\rho$ and $\omega$) exchanges are also allowed.
Thus, in future work we can study these systems by one boson
exchange (OBE) model by including all allowed light meson exchanges.
We also notice the studies by the local hidden gauge approach
\cite{Wu:2010jy}, where the vector light meson exchange is
considered, which shows that the vector light meson exchange can
provide an attraction. Thus, we can expect that the corresponding
$\Lambda$ value in the OBE model will become smaller than that in
OPE model, which means that the light vector meson exchange is
helpful to form these discussed molecular states.

\section*{Acknowledgments}

This project is supported by the National Natural Science Foundation
of China under Grants No. 11222547, No. 11175073, No. 11261130311
and 973 program. Xiang Liu is also supported by the National Youth
Top-notch Talent Support Program ("Thousands-of-Talents Scheme").


\begin{thebibliography}{99}

\bibitem{review}
 H.~X.~Chen, W.~Chen, X.~Liu and S.~L.~Zhu,
  The hidden-charm pentaquark and tetraquark states,
  arXiv:1601.02092 [hep-ph].


\bibitem{Aaij:2015tga}
  R.~Aaij {\it et al.} [LHCb Collaboration],
  Observation of $J\psi$ Resonances Consistent with Pentaquark States in $\Lambda_b^0\rightarrow J/\psi K^-p$ Decays,
  Phys.\ Rev.\ Lett.\  {\bf 115}, 072001 (2015)
  [arXiv:1507.03414 [hep-ex]].

\bibitem{Yang:2011wz}
  Z.~C.~Yang, Z.~F.~Sun, J.~He, X.~Liu and S.~L.~Zhu,
  The possible hidden-charm molecular baryons composed of anti-charmed meson and charmed baryon,
  Chin.\ Phys.\ C {\bf 36}, 6 (2012)
  [arXiv:1105.2901 [hep-ph]].

\bibitem{Wu:2010jy}
  J.~J.~Wu, R.~Molina, E.~Oset and B.~S.~Zou,
  Prediction of narrow $N^*$ and $\Lambda^*$ resonances with hidden charm above 4 GeV,
  Phys.\ Rev.\ Lett.\  {\bf 105}, 232001 (2010)
  [arXiv:1007.0573 [nucl-th]].

\bibitem{Wu:2010vk}
  J.~J.~Wu, R.~Molina, E.~Oset and B.~S.~Zou,
  Dynamically generated $N^{*}$ and $\Lambda^*$ resonances in the hidden charm sector around 4.3 GeV,
  Phys.\ Rev.\ C {\bf 84}, 015202 (2011)
  [arXiv:1011.2399 [nucl-th]].

\bibitem{Wang:2011rga}
  W.~L.~Wang, F.~Huang, Z.~Y.~Zhang and B.~S.~Zou,
  $\Sigma_c \bar{D}$ and $\Lambda_c \bar{D}$ states in a chiral quark model,
  Phys.\ Rev.\ C {\bf 84}, 015203 (2011)
  [arXiv:1101.0453 [nucl-th]].

\bibitem{Garcia-Recio:2013gaa}
  C.~Garcia-Recio, J.~Nieves, O.~Romanets, L.~L.~Salcedo and L.~Tolos,
  Hidden charm N and Delta resonances with heavy-quark symmetry,
  Phys.\ Rev.\ D {\bf 87}, 074034 (2013)
  [arXiv:1302.6938 [hep-ph]].

\bibitem{Xiao:2013yca}
  C.~W.~Xiao, J.~Nieves and E.~Oset,
  Combining heavy quark spin and local hidden gauge symmetries in the dynamical generation of hidden charm baryons,
  Phys.\ Rev.\ D {\bf 88}, 056012 (2013)
  [arXiv:1304.5368 [hep-ph]].

\bibitem{Yuan:2012wz}
  S.~G.~Yuan, K.~W.~Wei, J.~He, H.~S.~Xu and B.~S.~Zou,
  Study of $qqqc\bar{c}$ five quark system with three kinds of quark-quark hyperfine interaction,
  Eur.\ Phys.\ J.\ A {\bf 48}, 61 (2012)
  [arXiv:1201.0807 [nucl-th]].


\bibitem{Uchino:2015uha}
  T.~Uchino, W.~H.~Liang and E.~Oset,
  Baryon states with hidden charm in the extended local hidden gauge approach,
  arXiv:1504.05726 [hep-ph].

\bibitem{Chen:2015loa}
  R.~Chen, X.~Liu, X.~Q.~Li and S.~L.~Zhu,
  Identifying exotic hidden-charm pentaquarks,
  Phys.\ Rev.\ Lett.\  {\bf 115}, 132002 (2015)
  [arXiv:1507.03704 [hep-ph]].

\bibitem{Chen:2015moa}
  H.~X.~Chen, W.~Chen, X.~Liu, T.~G.~Steele and S.~L.~Zhu,
  Towards exotic hidden-charm pentaquarks in QCD,
  Phys.\ Rev.\ Lett.\  {\bf 115}, 172001 (2015)
  [arXiv:1507.03717 [hep-ph]].

\bibitem{Karliner:2015ina}
  M.~Karliner and J.~L.~Rosner,
  New Exotic Meson and Baryon Resonances from Doubly-Heavy Hadronic Molecules,
  Phys.\ Rev.\ Lett.\  {\bf 115}, no. 12, 122001 (2015)
  [arXiv:1506.06386 [hep-ph]].

\bibitem{Roca:2015dva}
  L.~Roca, J.~Nieves and E.~Oset,
  LHCb pentaquark as a $\bar{D}^*\Sigma_c-\bar{D}^*\Sigma_c^*$ molecular state,
  Phys.\ Rev.\ D {\bf 92}, 094003 (2015)
  [arXiv:1507.04249 [hep-ph]].

\bibitem{Mironov:2015ica}
  A.~Mironov and A.~Morozov,
  Is the pentaquark doublet a hadronic molecule?,
  JETP Lett.\  {\bf 102}, 271 (2015)
  [arXiv:1507.04694 [hep-ph]].

\bibitem{He:2015cea}
  J.~He,
  The $\bar{D}\Sigma^*_c$ and $\bar{D}^*\Sigma_c$ interactions and the LHCb hidden-charmed pentaquarks,
  arXiv:1507.05200 [hep-ph].

\bibitem{Meissner:2015mza}
  U.~G.~Meissner and J.~A.~Oller,
  Testing the $\chi_{c1}\, p$ composite nature of the $P_c(4450)$,
  Phys.\ Lett.\ B {\bf 751}, 59 (2015)
  [arXiv:1507.07478 [hep-ph]].

\bibitem{Burns:2015dwa}
  T.~J.~Burns,
  Phenomenology of $P_c(4380)^+$, $P_c(4450)^+$ and related states,
  arXiv:1509.02460 [hep-ph].

\bibitem{Huang:2015uda}
  H.~Huang, C.~Deng, J.~Ping and F.~Wang,
  Possible pentaquarks with heavy quarks,
  arXiv:1510.04648 [hep-ph].

\bibitem{Maiani:2015vwa}
  L.~Maiani, A.~D.~Polosa and V.~Riquer,
  The New Pentaquarks in the Diquark Model,
  Phys.\ Lett.\ B {\bf 749}, 289 (2015)
  [arXiv:1507.04980 [hep-ph]].

\bibitem{Anisovich:2015cia}
  V.~V.~Anisovich, M.~A.~Matveev, J.~Nyiri, A.~V.~Sarantsev and A.~N.~Semenova,
  Pentaquarks and resonances in the $pJ/\psi$ spectrum,
  arXiv:1507.07652 [hep-ph].

\bibitem{Li:2015gta}
  G.~N.~Li, M.~He and X.~G.~He,
  Some Predictions of Diquark Model for Hidden Charm Pentaquark Discovered at the LHCb,
  arXiv:1507.08252 [hep-ph].

\bibitem{Ghosh:2015ksa}
  R.~Ghosh, A.~Bhattacharya and B.~Chakrabarti,
  The masses of $P_{c}^{*}(4380)$ and $P_{c}^{*}(4450)$ in the quasi particle diquark model,
  arXiv:1508.00356 [hep-ph].

\bibitem{Wang:2015epa}
  Z.~G.~Wang,
  Analysis of the $P_c(4380)$ and $P_c(4450)$ as pentaquark states in the diquark model with QCD sum rules,
  arXiv:1508.01468 [hep-ph].

\bibitem{Anisovich:2015zqa}
  V.~V.~Anisovich, M.~A.~Matveev, J.~Nyiri, A.~V.~Sarantsev and A.~N.~Semenova,
  Non-strange and strange pentaquarks with hidden charm,
  Int.\ J.\ Mod.\ Phys.\ A {\bf 30}, 1550190 (2015)
  [arXiv:1509.04898 [hep-ph]].

\bibitem{Lebed:2015tna}
  R.~F.~Lebed,
  The Pentaquark Candidates in the Dynamical Diquark Picture,
  Phys.\ Lett.\ B {\bf 749}, 454 (2015)
  [arXiv:1507.05867 [hep-ph]].

\bibitem{Zhu:2015bba}
  R.~Zhu and C.~F.~Qiao,
  Novel Pentaquarks from Diquark-Triquark Model,
  arXiv:1510.08693 [hep-ph].

\bibitem{Guo:2015umn}
  F.~K.~Guo, U.~G.~Meissner, W.~Wang and Z.~Yang,
  Phys.\ Rev.\ D {\bf 92}, 071502 (2015)
  [arXiv:1507.04950 [hep-ph]].

\bibitem{Liu:2015fea}
  X.~H.~Liu, Q.~Wang and Q.~Zhao,
  Understanding the newly observed heavy pentaquark candidates,
  arXiv:1507.05359 [hep-ph].

\bibitem{Mikhasenko:2015vca}
  M.~Mikhasenko,
  A triangle singularity and the LHCb pentaquarks,
  arXiv:1507.06552 [hep-ph].

\bibitem{Scoccola:2015nia}
  N.~N.~Scoccola, D.~O.~Riska and M.~Rho,
  Pentaquark candidates P$_c^+$(4380) and P$_c^+$(4450) within the soliton picture of baryons,
  Phys.\ Rev.\ D {\bf 92}, 051501 (2015)
  [arXiv:1508.01172 [hep-ph]].


\bibitem{Yan:1992gz}
  T.~M.~Yan, H.~Y.~Cheng, C.~Y.~Cheung, G.~L.~Lin, Y.~C.~Lin and H.~L.~Yu,
  Heavy quark symmetry and chiral dynamics,
  Phys.\ Rev.\  D {\bf 46}, 1148 (1992)
  [Erratum-ibid.\  D {\bf 55}, 5851 (1997)].

\bibitem{Wise:1992hn}
  M.~B.~Wise,
  Chiral perturbation theory for hadrons containing a heavy quark,
  Phys.\ Rev.\  D {\bf 45}, 2188 (1992).

\bibitem{Burdman:1992gh}
  G.~Burdman and J.~F.~Donoghue,
  Union of chiral and heavy quark symmetries,
  Phys.\ Lett.\  B {\bf 280}, 287 (1992).

\bibitem{Casalbuoni:1996pg}
  R.~Casalbuoni, A.~Deandrea, N.~Di Bartolomeo, R.~Gatto, F.~Feruglio and G.~Nardulli,
  Phenomenology of heavy meson chiral Lagrangians,
  Phys.\ Rept.\  {\bf 281}, 145 (1997)
  [arXiv:hep-ph/9605342].

\bibitem{Falk:1992cx}
  A.~F.~Falk and M.~E.~Luke,
  Strong decays of excited heavy mesons in chiral perturbation theory,
  Phys.\ Lett.\  B {\bf 292}, 119 (1992)
  [arXiv:hep-ph/9206241].

\bibitem{Liu:2011xc}
  Y.~R.~Liu and M.~Oka,
  $\Lambda_c N$ bound states revisited,
  Phys.\ Rev.\ D {\bf 85}, 014015 (2012)
  [arXiv:1103.4624 [hep-ph]].


\bibitem{Liu:2008xz}
  X.~Liu, Y.~-R.~Liu, W.~-Z.~Deng and S.~-L.~Zhu,
  $Z^+(4430)$ as a $D_1'\bar{D}^*$, $(D_1\bar{D}^*)$ molecular state,
  Phys.\ Rev.\ D {\bf 77}, 094015 (2008)
  [arXiv:0803.1295 [hep-ph]].

\bibitem{Chen:2011cj}
  D.~Y.~Chen, X.~Liu and T.~Matsuki,
  Two Charged Strangeonium-Like Structures Observable in the $Y(2175) \to \phi(1020)\pi^{+} \pi^{-}$ Process,
  Eur.\ Phys.\ J.\ C {\bf 72}, 2008 (2012)
  [arXiv:1112.3773 [hep-ph]].

\bibitem{Oh:2000qr}
  Y.~s.~Oh, T.~Song and S.~H.~Lee,
  $J/\psi$ absorption by $\pi$ and $\rho$ mesons in meson exchange model with anomalous parity interactions,
  Phys.\ Rev.\ C {\bf 63}, 034901 (2001)
  [nucl-th/0010064].

\bibitem{Abrashkevichn:1995cj}
A. G. Abrashkevich, D. G. Abrashkevich, M. S. Kaschiev, I. V.
Puzynin, FESSDE, a Program for the Finite-element Solution of the
Coupled-channel Schr$\ddot{\text{o}}$dinger Equation Using
High-order Accuracy Approximations, Comput. Phys. Commun. 85, 65
(1995).

\bibitem{Abrashkevich:1998cj}
A. G. Abrashkevich, D. G. Abrashkevich, M. S. Kaschiev, I. V.
Puzynin, FESSDE 2.2: A New Version of a Program for the
Finite-element Solution of the Coupled-channel
Schr$\ddot{\text{o}}$dinger Equation Using High-order Accuracy
Approximations, Comput. Phys. Commun. 115, 90 (1998).

\bibitem{Yukawa:1935xg}
  H.~Yukawa,
  On the Interaction of Elementary Particles I,
  Proc.\ Phys.\ Math.\ Soc.\ Jap.\  {\bf 17}, 48 (1935)
  [Prog.\ Theor.\ Phys.\ Suppl.\  {\bf 1}, 1].

\bibitem{Machleidt:1989tm}
  R.~Machleidt,
  The Meson theory of nuclear forces and nuclear structure,
  Adv.\ Nucl.\ Phys.\  {\bf 19}, 189 (1989).

\bibitem{Machleidt:1987hj}
  R.~Machleidt, K.~Holinde and C.~Elster,
  The Bonn Meson Exchange Model for the Nucleon Nucleon Interaction,
  Phys.\ Rept.\  {\bf 149}, 1 (1987).

\bibitem{Epelbaum:2008ga}
  E.~Epelbaum, H.~W.~Hammer and U.~G.~Meissner,
  Modern Theory of Nuclear Forces,
  Rev.\ Mod.\ Phys.\  {\bf 81}, 1773 (2009)
  [arXiv:0811.1338 [nucl-th]].


\bibitem{Aubert:2007dt}
  B.~Aubert {\it et al.} [BaBar Collaboration],
  A Study of Excited Charm-Strange Baryons with Evidence for new Baryons $\Xi_c(3055)^+$ and $\Xi_c(3123)^+$,
  Phys.\ Rev.\ D {\bf 77}, 012002 (2008)
  [arXiv:0710.5763 [hep-ex]].

\bibitem{Ebert:2007nw}
  D.~Ebert, R.~N.~Faustov and V.~O.~Galkin,
  Masses of excited heavy baryons in the relativistic quark-diquark picture,
  Phys.\ Lett.\ B {\bf 659}, 612 (2008)
  [arXiv:0705.2957 [hep-ph]].



\bibitem{Wu:2010rv}
  J.~J.~Wu and B.~S.~Zou,
  Prediction of super-heavy $N^*$ and $\Lambda^*$ resonances with hidden beauty,
  Phys.\ Lett.\ B {\bf 709}, 70 (2012)
  [arXiv:1011.5743 [hep-ph]].






\end{thebibliography}
\end{document}